\documentclass[12pt]{article}
\pdfoutput=1

\usepackage[a4paper,text={16.8cm,22.4cm}]{geometry}
\usepackage{amsmath,amsfonts,braket,slashed,amssymb,bm,psfrag,graphicx,color,dsfont,euscript,ctable,bbm}
\usepackage[small,labelfont=bf]{caption}
\RequirePackage[sort&compress,square,comma,numbers]{natbib}
\allowdisplaybreaks
\newcommand{\spac}{{\hspace{0.3mm}}}

\begin{document}

\begin{titlepage}

\begin{flushright}
\normalsize
IPPP/20/69\\
MITP/20-070\\
SISSA 30/2020/FISI\\
ZH-TH-47/20\\
December 22, 2020
\end{flushright}
% arXiv:2012.12272
% v1: December 22, 2020
% v1a: December 31, 2020 (submission to JHEP)
% v2: February 10, 2021 (resubmission to JHEP)
% v2a: March 30, 2021 (published JHEP version including proof corrections)

\vspace{1mm}
\begin{center}
\Large\bf
The Low-Energy Effective Theory of Axions and ALPs
\end{center}

\vspace{1mm}
\begin{center}
Martin Bauer$^a$, Matthias Neubert$^{b,c,d}$, Sophie Renner$^e$, Marvin Schnubel$^b$\\ 
and Andrea Thamm$^f$\\
\vspace{5mm} 
{\small\sl ${}^a$Institute for Particle Physics Phenomenology, Department of Physics\\
Durham University, Durham, DH1 3LE, UK\\[2mm]
${}^b$PRISMA$+$ Cluster of Excellence \& Mainz Institute for Theoretical Physics\\
Johannes Gutenberg University, 55099 Mainz, Germany\\[2mm]
${}^c$Physik-Institut, Universit\"at Z\"urich, CH-8057, Switzerland\\[2mm]
${}^d$Department of Physics \& LEPP, Cornell University, Ithaca, NY 14853, U.S.A.\\[2mm]
${}^e$SISSA International School for Advanced Studies, Via Bonomea 265, 34136, Trieste, Italy\\
INFN, Sezione di Trieste, Via Bonomea 265, 34136 Trieste, Italy\\[2mm]
${}^f$School of Physics, The University of Melbourne, Victoria 3010, Australia}
\end{center}

\vspace{5mm}
\begin{abstract}
Axions and axion-like particles (ALPs) are well-motivated low-energy relics of high-energy extensions of the Standard Model, which interact with the known particles through higher-dimensional operators suppressed by the mass scale $\Lambda$ of the new-physics sector. Starting from the most general dimension-5 interactions, we discuss in detail the evolution of the ALP couplings from the new-physics scale to energies at and below the scale of electroweak symmetry breaking. We derive the relevant anomalous dimensions at two-loop order in gauge couplings and one-loop order in Yukawa interactions, carefully considering the treatment of a redundant operator involving an ALP coupling to the Higgs current. We account for one-loop (and partially two-loop) matching contributions at the weak scale, including in particular flavor-changing effects. The relations between different equivalent forms of the effective Lagrangian are discussed in detail. We also construct the effective chiral Lagrangian for an ALP interacting with photons and light pseudoscalar mesons, pointing out important differences with the corresponding Lagrangian for the QCD axion. 
\end{abstract}

\end{titlepage}

\tableofcontents
\vspace{1.1cm}

\section{Introduction}

Axions and axion-like particles (ALPs) are pseudo Nambu--Goldstone bosons, which appear in the spontaneous breaking of a global symmetry and are well motivated new-physics relics in a variety of explicit extensions of the Standard Model (SM) of elementary-particle physics. Their name is derived from the QCD axion, which was introduced by Peccei, Quinn and others to address the strong CP problem \cite{Peccei:1977hh,Peccei:1977ur,Weinberg:1977ma,Wilczek:1977pj}. While several explicit models of QCD axions \cite{Kim:1979if,Shifman:1979if,Dine:1981rt,Zhitnitsky:1980tq} predict a rather strict relation between the axion mass and decay constant, it was realized early on that it is possible to obtain solutions to the strong CP problem with heavier ALPs \cite{Rubakov:1997vp}. Furthermore, supersymmetric and composite-Higgs models can naturally feature light pseudoscalar particles. For example, the R-axion is the pseudo Nambu--Goldstone boson of the R-symmetry breaking in low-energy supersymmetry \cite{Bellazzini:2017neg}, while non-minimal coset structures in models of compositeness predict pseudo Nambu--Goldstone bosons in addition to the Higgs boson \cite{Ferretti:2013kya}. These models provide ample motivation to search for light ALPs, in particular those with masses in the range between an MeV and tens of GeV, whose couplings are not tightly constrained by existing cosmological \cite{Cadamuro:2011fd,Millea:2015qra}, astrophysical \cite{Payez:2014xsa,Jaeckel:2017tud} and collider bounds \cite{Chen:2010su,Mimasu:2014nea,Jaeckel:2015jla,Knapen:2016moh,Brivio:2017ije,Bauer:2017nlg,Bauer:2017ris,Knapen:2017ebd,Mariotti:2017vtv,Craig:2018kne,Bauer:2018uxu,Aloni:2018vki,Alonso-Alvarez:2018irt,Aloni:2019ruo,Baldenegro:2019whq,Gavela:2019cmq,Coelho:2020saz}.

The results of this work apply equally to the cases of the QCD axion and of a more general ALP, and from now on we use the term ALP to represent both options. We use a model-independent approach to connect the ALP couplings to SM particles, which can be probed in low-energy experiments, with the couplings at the fundamental new-physics scale $\Lambda$, which we assume to be far above the scale of electroweak symmetry breaking. The leading-order interactions with SM fields can be parameterized in terms of the Wilson coefficients of dimension-5 operators suppressed by $1/\Lambda$, and hence a heavy new-physics sector corresponds to weak ALP couplings. Starting from the most general effective Lagrangian at dimension-5 order, we calculate the effects of renormalization-group (RG) evolution from the new-physics scale down to the scale of electroweak symmetry breaking and below, systematically including all contributions to the anomalous dimensions arising at two-loop order in gauge couplings and one-loop order in Yukawa interactions. The effects of a redundant operator, in which the ALP couples to the Higgs current, are carefully taken into account. We also calculate the complete one-loop matching contributions at the weak scale, which arise when the top quark, the Higgs boson and the $W$ and $Z$ bosons are integrated out. If the underlying global symmetry is flavor-dependent, the ALP couplings to quarks or leptons can have a non-trivial flavor structure at the scale $\Lambda$ \cite{Davidson:1981zd,Ema:2016ops,Calibbi:2016hwq}. But even if the underlying global symmetry is flavor-universal, flavor-violating ALP couplings are inevitably induced radiatively. This opens up the possibility to search for ALPs in rare, flavor-changing decays of mesons and leptons, which could provide information about the structure of a new-physics sector otherwise out of reach of direct searches. We illustrate the numerical effects of RG evolution and weak-scale matching for different values of the new-physics scale $\Lambda$. Our study of these effects goes significantly beyond existing studies in the literature, and it is relevant for the case of the QCD axion, too. We also discuss the relations between several equivalent forms of the effective ALP Lagrangian, which differ in the form of the ALP--fermion interactions. Finally, we discuss the matching of the effective Lagrangian at low energies onto a chiral effective Lagrangian describing the couplings of a light ALP to photons and light pseudoscalar mesons, carefully taking into account the presence of a non-zero ALP mass in the effective theory, which gives rise to several important effects.

The results of this work form the basis for precise phenomenological analyses of the physics of a light ALP or axion, connecting low-energy observables in a systematic and accurate way with the couplings of the underlying ultra-violet (UV) complete theory.

\section{ALP couplings to the SM}
\label{sec:2}

We consider a gauge-singlet, pseudoscalar resonance $a$, whose couplings to SM fields are, at the classical level, protected by an approximate shift symmetry $a\to a+c$, broken softly by the mass term $m_{a,0}^2$. Such a coupling structure arises, for example, if the particle $a$ can be identified with the phase of a complex scalar field. 

\subsection{Choice of the operator basis} 

The most general effective Lagrangian for this particle including operators of up to dimension~5 reads \cite{Georgi:1986df}\footnote{The ALP couplings to fermions and gauge bosons in (\ref{Leff}) are related to the analogous couplings introduced in \cite{Bauer:2017ris} by $f=\Lambda/(4\pi)$, $\bm{c}_{F}=\bm{C_{F}}/(4\pi)$ and $c_{VV}=4\pi\,C_{VV}$ with $V=G,W,B$.}  
\begin{equation}\label{Leff}
\begin{aligned}
   {\cal L}_{\rm eff}^{D\le 5}
   &= \frac12 \left( \partial_\mu a\right)\!\left( \partial^\mu a\right) - \frac{m_{a,0}^2}{2}\,a^2
    + \frac{\partial^\mu a}{f}\,\sum_F\,\bar\psi_F\spac\bm{c}_F\spac\gamma_\mu\spac\psi_F \\
   &\quad\mbox{}+ c_{GG}\,\frac{\alpha_s}{4\pi}\,\frac{a}{f}\,G_{\mu\nu}^a\,\tilde G^{\mu\nu,a}
    + c_{WW}\,\frac{\alpha_2}{4\pi}\,\frac{a}{f}\,W_{\mu\nu}^A\,\tilde W^{\mu\nu,A}
    + c_{BB}\,\frac{\alpha_1}{4\pi}\,\frac{a}{f}\,B_{\mu\nu}\,\tilde B^{\mu\nu} \spac .
\end{aligned}
\end{equation}
Here $G_{\mu\nu}^a$, $W_{\mu\nu}^A$ and $B_{\mu\nu}$ are the field-strength tensors of $SU(3)_c$, $SU(2)_L$ and $U(1)_Y$, and $\alpha_s=g_s^2/(4\pi)$, $\alpha_2=g^2/(4\pi)$ and $\alpha_1=g^{\prime\,2}/(4\pi)$ denote the corresponding coupling parameters. $\tilde B^{\mu\nu}=\frac12\epsilon^{\mu\nu\alpha\beta} B_{\alpha\beta}$ etc.\ (with $\epsilon^{0123}=1$) are the dual field-strength tensors. The sum in the first line extends over the chiral fermion multiplets $F$ of the SM. The quantities $\bm{c}_F$ are hermitian matrices in generation space. For the couplings of $a$ to the $U(1)_Y$ and $SU(2)_L$ gauge fields, the additional terms arising from a constant shift $a\to a+c$ of the ALP field can be removed by field redefinitions. The coupling to QCD gauge fields is not invariant under a continuous shift transformation because of instanton effects, which however preserve a discrete version of the shift symmetry, under which $a\to a+n\spac\pi f/c_{GG}$ with integer $n$ \cite{Weinberg:1977ma,Wilczek:1977pj}. Above we have indicated the suppression of the dimension-5 operators with the ALP decay constant $f$, which is related to the relevant new-physics scale by $\Lambda=4\pi f$. This is the characteristic scale of global symmetry breaking, assumed to be far above the weak scale. It is then a good approximation to neglect contributions from higher-dimensional operators, which are suppressed by higher powers of $1/f$.\footnote{In the literature on QCD axions $f$ is often eliminated in favor of the axion decay constant $f_a$, defined such that $1/f_a\equiv -2c_{GG}/f$. The parameter $1/f_a$ then determines the strength of the axion--gluon coupling.} 
Since our effective theory only contains the SM particles and the ALP as degrees of freedom, it would need to be modified in scenarios with a new-physics sector between the weak scale and the scale of global symmetry breaking ($v<M_{\rm NP}<4\pi f$). Even in this case, the effective Lagrangian (\ref{Leff}) offers a model-independent description of the physics below the intermediate scale $M_{\rm NP}$.

The physical ALP mass is given by the sum of the explicit soft breaking term $m_{a,0}^2$ and the contribution to the mass generated by non-perturbative QCD dynamics \cite{Bardeen:1978nq,Shifman:1979if,DiVecchia:1980yfw}, such that at lowest order in chiral perturbation theory 
\begin{equation}\label{massrela}
   m_a^2 = m_{a,0}^2 \left[ 1 + {\cal O}\bigg(\frac{f_\pi^2}{f^2}\bigg) \right]
    + c_{GG}^2\,\frac{f_\pi^2\,m_\pi^2}{f^2}\,\frac{2 m_u m_d}{(m_u + m_d)^2} \,,
\end{equation}
where $f_\pi\simeq 130$\,MeV is the pion decay constant. The correction to the first term in this relation will be discussed in Section~\ref{sec:chiPT}. Whereas for the classical QCD axion (with $m_{a,0}^2=0$) there is a strict relation between the mass and the coupling to gluons, the presence of the additional contribution $m_{a,0}^2$ allows for heavier ALPs, which however are still naturally much lighter than the scale $f$ as long as the ALP is a pseudo Nambu--Goldstone boson and the shift symmetry is effective. It is possible to generate this additional contribution dynamically using non-abelian extensions of the SM, in which additional instanton contributions arise \cite{Holdom:1982ex,Holdom:1985vx,Dine:1986bg,Flynn:1987rs,Choi:1988sy,Choi:1998ep,Rubakov:1997vp,Berezhiani:2000gh,Hall:2014vga,Hook:2014cda,Fukuda:2015ana,Dimopoulos:2016lvn,Agrawal:2017ksf,Gaillard:2018xgk}, or using the recently proposed mechanism of axion kinetic misalignment, in which the axion shift symmetry is explicitly broken in the early universe \cite{Co:2019jts}. It is thus possible to generate an ALP mass significantly larger than the contribution from QCD instantons while preserving the Peccei--Quinn solution of the strong CP problem. 

The ALP couplings $\bm{c}_F$ to the SM fermions can, in principle, have a non-trivial structure in generation space, thereby giving rise to flavor-changing neutral current interactions mediated by ALP exchange. The phenomenological constraints on such couplings are very strong, especially for light ALPs, which can be produced in the decays of kaons or $B$ mesons \cite{Batell:2009jf,Izaguirre:2016dfi,Choi:2017gpf,CidVidal:2018blh,Gavela:2019wzg,MartinCamalich:2020dfe,inprep}, and which can give sizable contributions to flavor-changing transitions in the lepton sector \cite{Bauer:2019gfk,Cornella:2019uxs,Calibbi:2020jvd} and to electric dipole moments \cite{Marciano:2016yhf,DiLuzio:2020oah}. In extensions of the SM in which the new-physics scale $\Lambda=4\pi f$ is not very far above the TeV scale, the coupling matrices $\bm{c}_F$ must have a hierarchical structure in order to be consistent with these constraints. From the point of view of model building, such a structure can be ensured by imposing the principle of minimal flavor violation \cite{DAmbrosio:2002vsn}. Under this hypothesis, the matrices $\bm{c}_Q$ and $\bm{c}_q$ in the quark sector can be expanded as 
\begin{equation}\label{MFV1}
\begin{aligned}
   \bm{c}_Q &= c_0^Q\,\mathbbm{1} + \epsilon \left( c_1^Q\,\bm{Y}_u\spac\bm{Y}_u^\dagger
    + c_2^Q\,\bm{Y}_d\spac\bm{Y}_d^\dagger \right) + {\cal O}(\epsilon^2) \,, \\
   \bm{c}_u &= c_0^u\,\mathbbm{1} + \epsilon\,c_1^u\,\bm{Y}_u^\dagger\spac\bm{Y}_u 
    + \epsilon^2 \left[ c_2^u\,\big( \bm{Y}_u^\dagger\spac\bm{Y}_u \big)^2
    + c_3^u\,\bm{Y}_u^\dagger\spac\bm{Y}_d\spac\bm{Y}_d^\dagger\spac\bm{Y}_u \right] 
    + {\cal O}(\epsilon^3) \,, \\
   \bm{c}_d &= c_0^d\,\mathbbm{1} + \epsilon\,c_1^d\,\bm{Y}_d^\dagger\spac\bm{Y}_d 
    + \epsilon^2 \left[ c_2^d\,\big( \bm{Y}_d^\dagger\spac\bm{Y}_d \big)^2
    + c_3^d\,\bm{Y}_d^\dagger\spac\bm{Y}_u\spac\bm{Y}_u^\dagger\spac\bm{Y}_d \right] 
    + {\cal O}(\epsilon^3) \,,
\end{aligned}
\end{equation}
where $\epsilon$ counts the order in the spurion expansion. Analogous expressions apply in the lepton sector. The phenomenological implications of these results will be discussed later.

\subsection{A redundant operator}

The form of the effective Lagrangian (\ref{Leff}) is not unique. At dimension-5 order one can also write down an ALP coupling to the Higgs doublet $\phi$, given by
\begin{equation}\label{Higgscoupling}
   {\cal L}_{\rm eff}^{D\le 5} \supset c_\phi\,O_\phi
   = c_\phi\,\frac{\partial^\mu a}{f} \left( \phi^\dagger\spac iD_\mu\spac\phi + \mbox{h.c.} \right) .
\end{equation}
The operator $O_\phi$ is redundant, however, because it can be reduced to the fermionic operators in (\ref{Leff}) using the field equations for the Higgs doublet and the SM fermions \cite{Georgi:1986df}. Indeed, the field redefinitions $\phi\to e^{ic_\phi\spac a/f}\,\phi$ and $F\to e^{-i\beta_F\spac c_\phi\spac a/f}\,F$ for all chiral fermion multiplets $F$ of the SM, subject to the conditions 
\begin{equation}\label{betacond}
   \beta_u - \beta_Q = - 1 \,, \qquad \beta_d - \beta_Q = 1 \,, \qquad
   \beta_e - \beta_L = 1 \,, \qquad 3\beta_Q + \beta_L = 0 \,, 
\end{equation}
eliminate the term $c_\phi\,O_\phi$ from the Lagrangian at the expense of shifting the flavor matrices $\bm{c}_F$ by 
\begin{equation}\label{cfshifts}
   \bm{c}_F \to \bm{c}_F + \beta_F\,c_\phi\,\mathbbm{1} \,.
\end{equation}
The first three relations in (\ref{betacond}) ensure that the SM Yukawa interactions are invariant under the field redefinitions. The fourth relation guarantees that the combination of fermion currents induced by the field redefinitions is anomaly free, and hence no additional contributions to the coefficients of the operators in (\ref{Leff}) involving the gauge fields are generated.
 
The conditions (\ref{betacond}) define a one-parameter class of field redefinitions, which one can use to eliminate the operator $O_\phi$ from the effective Lagrangian. One particular solution is given by the choice $\beta_u=-1$, $\beta_d=\beta_e=1$ and $\beta_Q=\beta_L=0$, which was adopted in \cite{Bauer:2016ydr,Bauer:2016zfj} and eliminates $O_\phi$ in favor of a linear combination of operators involving right-handed quark currents. A different solution consists of the choice $\beta_F=-2\spac{\cal Y}_F$, where ${\cal Y}_F$ denotes the hypercharge of the fermion multiplet $F$ \cite{Georgi:1986df,MartinCamalich:2020dfe}. In general, the derivative couplings of the ALP are only defined modulo generators of exact global symmetries of the SM, which include baryon and lepton number. We will see later that physical quantities are independent of the particular choice of $\beta_F$ values as long as the conditions (\ref{betacond}) are satisfied.

It follows from this discussion that the redundant operator $O_\phi$ can be re-expressed in the form
\begin{equation}
   O_\phi = {\cal O}_\phi + \sum_F\,\beta_F\,O_F \,, 
    \qquad \text{with} \quad 
   O_F = \frac{\partial^\mu a}{f}\,\bar\psi_F^i\spac\gamma_\mu\spac\psi_F^i \,,
\end{equation}
where a sum over the generation index $i$ is implied, and the new operator ${\cal O}_\phi$ vanishes by the equations of motion. It is a well-known fact that such operators do not need to be included in the renormalization of the basis operators in an effective field theory \cite{Politzer:1980me,Georgi:1991ch}. Hence, it is consistent to leave out the operator ${\cal O}_\phi$ from the effective Lagrangian (\ref{Leff}).  As we will see in Section~\ref{sec:RGE}, the original operator $O_\phi$ is needed as a counterterm to absorb some UV divergences of loop diagrams involving the fermionic operators $O_F$. The correct treatment then consists of projecting $O_\phi$ back onto our basis using the replacement rule \cite{Buchler:2003vw,Elias-Miro:2013mua,Jenkins:2013zja}
\begin{equation}\label{Ophireplace}
   O_\phi \to \sum_F\,\beta_F\,O_F \,.
\end{equation}

\subsection{Equivalent forms of the effective Lagrangian}

Another important freedom in writing down the effective Lagrangian concerns the structure of the ALP couplings to fermions. One can integrate by parts in the third term in (\ref{Leff}) and use the SM equations of motion along with the well-known equation for the axial anomaly to put the effective Lagrangian in the alternative form
\begin{equation}\label{Leffalt}
\begin{aligned}
   {\cal L}_{\rm eff}^{D\le 5}
   &= \frac12 \left( \partial_\mu a\right)\!\left( \partial^\mu a\right) - \frac{m_{a,0}^2}{2}\,a^2
    - \frac{a}{f} \left( \bar Q\spac\phi\spac\tilde{\bm{Y}}_d\,d_R 
    + \bar Q\spac\tilde\phi\spac\tilde{\bm{Y}}_u\spac u_R 
    + \bar L\spac\phi\spac\tilde{\bm{Y}}_e\spac e_R + \text{h.c.} \right) \\
   &\quad\mbox{}+ \tilde c_{GG}\,\frac{\alpha_s}{4\pi}\,\frac{a}{f}\,G_{\mu\nu}^a\,\tilde G^{\mu\nu,a}
    + \tilde c_{WW}\,\frac{\alpha_2}{4\pi}\,\frac{a}{f}\,W_{\mu\nu}^A\,\tilde W^{\mu\nu,A}
    + \tilde c_{BB}\,\frac{\alpha_1}{4\pi}\,\frac{a}{f}\,B_{\mu\nu}\,\tilde B^{\mu\nu} \spac ,
\end{aligned}
\end{equation}
where
\begin{equation}\label{rela1}
   \tilde{\bm{Y}}_d = i\hspace{0.3mm} \big( \bm{Y}_d\,\bm{c}_d - \bm{c}_Q \bm{Y}_d \big) \,, \qquad
   \tilde{\bm{Y}}_u = i\hspace{0.3mm} \big( \bm{Y}_u\,\bm{c}_u - \bm{c}_Q \bm{Y}_u \big) \,, \qquad
   \tilde{\bm{Y}}_e = i\hspace{0.3mm} \big( \bm{Y}_e\,\bm{c}_e - \bm{c}_L \bm{Y}_e \big) \,,
\end{equation}
and
\begin{equation}\label{rela2}
\begin{aligned}
   \tilde c_{GG} &= c_{GG} + T_F\,\text{Tr} \left( \bm{c}_u + \bm{c}_d - N_L\,\bm{c}_Q \right) , \\[1mm]
   \tilde c_{WW} &= c_{WW} - T_F\,\text{Tr} \left( N_c\,\bm{c}_Q + \bm{c}_L \right) , \\[-0.5mm]
   \tilde c_{BB} &= c_{BB} + \text{Tr}\,\Big[ N_c \left( {\cal Y}_u^2\,\bm{c}_u 
    + {\cal Y}_d^2\,\bm{c}_d - N_L\,{\cal Y}_Q^2\,\bm{c}_Q \right)
    + {\cal Y}_e^2\,\bm{c}_e - N_L\,{\cal Y}_L^2\,\bm{c}_L \Big] \,.
\end{aligned}
\end{equation}
Here the traces are over generation indices. $T_F=\frac12$ fixes the normalization of the $SU(N)$ group generators, $N_c=3$ is the number of colors, and $N_L=2$ denotes the number of weak isospin components. ${\cal Y}_Q=\frac16$, ${\cal Y}_u=\frac23$, ${\cal Y}_d=-\frac13$, ${\cal Y}_L=-\frac12$ and ${\cal Y}_e=-1$ denote the hypercharge quantum numbers of the SM quarks and leptons. The effective Lagrangians (\ref{Leff}) and (\ref{Leffalt}) are equivalent as long as these relations are taken into account. Note, however, that in (\ref{Leffalt}) there is no apparent reason for the complex matrices $\tilde{\bm{Y}}_f$ to have any particular structure. It is the shift symmetry encoded in the effective ALP Lagrangian (\ref{Leff}) that gives rise to the hierarchical structure of these matrices, which results from the appearance of the SM Yukawa matrices in (\ref{rela1}). This feature distinguishes an ALP from a generic pseudoscalar boson $a$. We thus prefer to take the Lagrangian (\ref{Leff}) as the starting point of our calculations. Nevertheless, we will see that the combinations $\tilde c_{VV}$ of ALP--boson and ALP--fermion couplings shown in (\ref{rela2}) play an important role in phenomenological applications of the effective Lagrangian and in the evolution of the ALP couplings from the new-physics scale $\Lambda$ down to lower energies.

\begin{figure}
\begin{center}
\includegraphics[height=2.3cm]{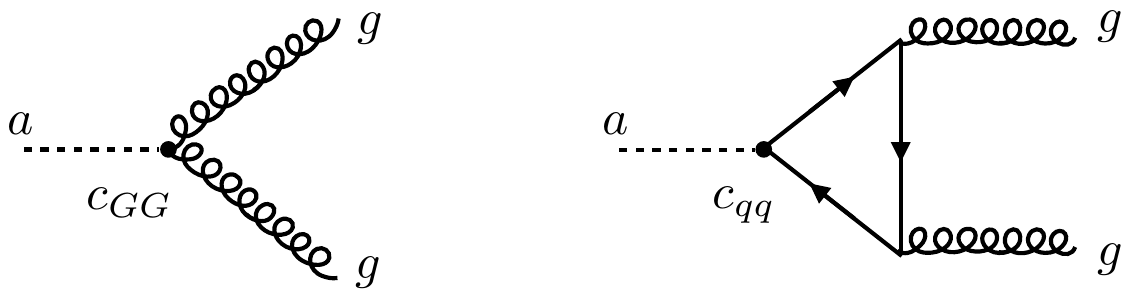}
\caption{\label{fig:agg} 
Contributions to the $a\to gg$ decay amplitude involving the ALP--gluon coupling (left) and the ALP couplings to quarks (right). The ALP is drawn as a dotted line. The black circles indicate vertices deriving from the dimension-5 operators in the effective Lagrangian (\ref{Leff}).} 
\vspace{-3mm}
\end{center}
\end{figure}

It is instructive to illustrate the equivalence of the effective Lagrangians (\ref{Leff}) and (\ref{Leffalt}) with a concrete example. Consider the decay of an ALP with mass $m_a\gg\Lambda_{\rm QCD}$ into two gluons, which manifest themselves as two jets in the final state. The relevant contributions to the decay amplitude are shown in Figure~\ref{fig:agg}. Calculating the decay rate at one-loop order in perturbation theory, taking into account radiative corrections calculated in \cite{Spira:1995rr}, one obtains \cite{Bauer:2017ris} 
\begin{equation}\label{aggrate}
   \Gamma(a\to gg)
   = \frac{\alpha_s^2(m_a)\,m_a^3}{8\pi^3 f^2} 
    \left[ 1 + \left( \frac{97}{4} - \frac{7 n_q}{6} \right) \frac{\alpha_s(m_a)}{\pi} \right]
    \left| C_{gg}^{\rm eff} \right|^2 .
\end{equation}
Here $n_q$ is the number of light quark flavors with mass below the ALP mass, and
\begin{equation}
   C_{gg}^{\rm eff}
   = c_{GG} + \frac12\spac\sum_q\spac c_{qq}(m_a)\,B_1\bigg(\frac{4m_q^2}{m_a^2}\bigg) \,,
\end{equation}
where
\begin{equation}\label{B1def}
   B_1(\tau) = 1 - \tau\spac f^2(\tau) \,, \qquad \mbox{with} \quad
   f(\tau) = \left\{ \begin{array}{ll} 
    \arcsin\frac{1}{\sqrt{\tau}} \,; &~ \tau\ge 1 \,, \\
    \frac{\pi}{2} + \frac{i}{2} \ln\frac{1+\sqrt{1-\tau}}{1-\sqrt{1-\tau}} \,; &~ \tau<1 \,.
    \end{array} \right. 
\end{equation} 
The sum runs over the six quark species of the SM. The parameters $c_{qq}(m_a)$ describe the flavor-diagonal ALP couplings to the quark mass eigenstates and will be defined later in (\ref{cffdef}). They are connected with the ALP--fermion couplings $\bm{c}_q$ and $\bm{c}_Q$ after these have been transformed into the mass basis of the SM quarks. The above result is obtained based on the effective Lagrangian (\ref{Leff}). If instead the calculations are starting from the alternative form of the effective Lagrangian shown in (\ref{Leffalt}), one finds
\begin{equation}
   C_{gg}^{\rm eff}
   = \tilde c_{GG} + \frac12\spac\sum_q\spac c_{qq}(m_a) \left[ B_1\bigg(\frac{4m_q^2}{m_a^2}\bigg) 
    - 1 \right] .
\end{equation}
The ``$-1$'' inside the bracket accounts for the difference in the fermion loop function, which is a consequence of the difference in the Feynman rules for the ALP--fermion vertices derived from the two Lagrangians. At the same time, the coefficient $\tilde c_{GG}$ differs from $c_{GG}$ by the terms shown in the first equation in (\ref{rela2}). Because of the trace, the difference between the two parameters is invariant under the unitary transformation to the mass basis, and one finds 
\begin{equation}
   \tilde c_{GG} = c_{GG} + \frac12\,\text{Tr} \left( \bm{c}_u + \bm{c}_d - 2\bm{c}_Q \right) 
   = c_{GG} + \frac12\,\sum_q\,c_{qq} \,.
\end{equation}
We thus find that the above two relations for $C_{gg}^{\rm eff}$ are indeed equivalent.

It is possible to work with a hybrid form of the effective ALP Lagrangian, in which the ALP--fermion interactions consist of both derivative terms, such as in (\ref{Leff}), and non-derivative terms, such as in (\ref{Higgscoupling}). This is useful, in particular, for low-energy applications in the context of the chiral effective Lagrangian. We will come back to this point in Section~\ref{sec:chiPT}.

\begin{figure}
\begin{center}
\includegraphics[height=2.0cm]{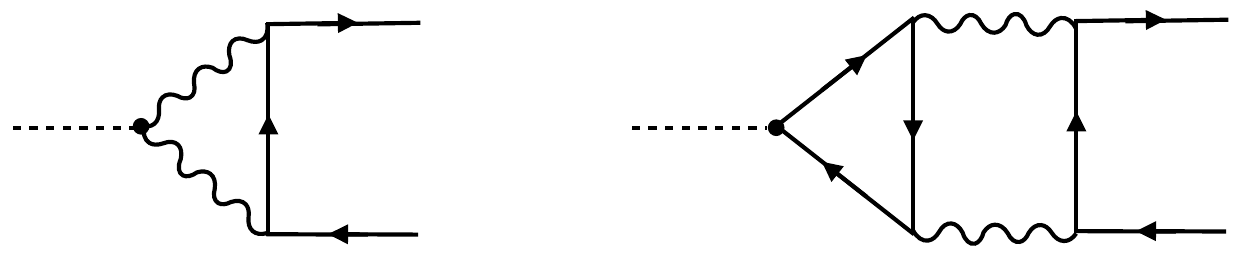}
\caption{\label{fig:2loop} 
Examples of one-loop and two-loop diagrams contributing at the same order in perturbation theory if $c_{VV}$ and $\bm{c}_F$ have similar magnitude.} 
\end{center}
\end{figure}

Our definitions of the ALP couplings in (\ref{Leff}) are such that the parameters $c_{VV}$ and $\bm{c}_F$ are expected to be of ${\cal O}(1)$ when one applies the counting rules of naive dimensional analysis \cite{Manohar:1983md,Luty:1997fk,Cohen:1997rt}. These rules imply, in particular, that the ALP--boson couplings $c_{VV}$ should be accompanied by a loop factor $\sim\alpha_i/(4\pi)$, as shown in (\ref{Leff}). However, one can conceive models in which these couplings are induced by loops involving a parametrically large number $N_f$ of new heavy fermions, such that $c_{VV}\propto N_f\gg 1$ can (at least partially) compensate for the loop suppression. In our analysis below, we account for this possibility by including the one-loop corrections proportional to the ALP--boson couplings in the RG equations for the ALP--fermion couplings, even though they provide two-loop contributions $\sim(\alpha_i/\pi)^2$ to these equations. A second rationale for this approach lies in the fact that in many concrete ALP models only certain ALP couplings are non-zero at the UV scale. Our treatment in the next section captures the leading contributions in each coupling irrespective of the relative magnitude of the ALP--boson and ALP--fermion couplings in the high-energy theory. We emphasize, however, that in cases where the coefficients $c_{VV}$ and $\bm{c}_F$ are of similar magnitude, one-loop diagrams involving the coefficients $c_{VV}$ have the same scaling as two-loop diagrams involving the coefficients $\bm{c}_F$, see Figure~\ref{fig:2loop}. For consistency, we thus include {\em all\/} two-loop contributions in the gauge couplings in the RG equations for the ALP--fermion couplings.

\section{Renormalization-group evolution to the weak scale}
\label{sec:RGE}

The effective Lagrangian (\ref{Leff}) is assumed to arise from integrating out some new heavy particles at a scale $\Lambda=4\pi f$ far above the weak scale. Assuming the ALP mass is small -- of order 100\,GeV or less -- we can evolve the Wilson coefficients and operators in the effective Lagrangian down to the scale of electroweak symmetry breaking by solving their RG equations. We now derive the explicit form of these equations, working consistently at two-loop order in gauge couplings and one-loop order in Yukawa interactions. These are the lowest orders at which these interactions contribute to the evolution equations for the ALP couplings. In models in which the boson couplings are enhanced over the fermion ones, the two-loop gauge contributions can give rise to the dominant evolution effects. Two-loop corrections in the Yukawa couplings, or mixed two-loop gauge--Yukawa contributions, are neglected in our approach. They would give rise to small multiplicative corrections of the fermion couplings, but they do not introduce new ALP coupling parameters on the right-hand side of the evolution equations. Thus, there is no scenario in which these neglected two-loop contributions could give rise to dominant effects. Some technical details of our derivations are relegated to Appendix~\ref{app:A}. The RG equations for the ALP couplings appearing in the alternative form of the effective Lagrangian in (\ref{Leffalt}) can be derived from the equations below in a straightforward way. They are discussed in Appendix~\ref{app:B}.

\subsection{Derivation of the RG evolution equations}

Pulling out one factor of $\alpha_i$ in the definitions of the ALP couplings to gauge fields in (\ref{Leff}) ensures that the Wilson coefficients $c_{VV}$ are scale independent (at least up to two-loop order in gauge couplings), i.e.\ 
\begin{equation}\label{cVVRGE}
   \frac{d}{d\ln\mu}\,c_{VV}(\mu) = 0 \,; \quad V = G,W,B \,.
\end{equation}
For the QCD coefficient $c_{GG}$ this follows from the explicit calculations performed in \cite{Chetyrkin:1998mw}, and an analogous statement holds for $c_{WW}$ and $c_{BB}$. This is different from the case of a scalar (CP-even) field coupled to two gauge fields, in which the corresponding couplings exhibit a non-trivial RG evolution starting at two-loop order \cite{Inami:1982xt,Grinstein:1988wz}. We have checked explicitly that the one-loop diagrams involving the scalar Higgs doublet do not give rise to a scale dependence of the coefficients $c_{WW}$ and $c_{BB}$ either. The contributions from these graphs are absorbed by the renormalization of the gauge couplings.

\begin{figure}
\begin{center}
\includegraphics[height=4.5cm]{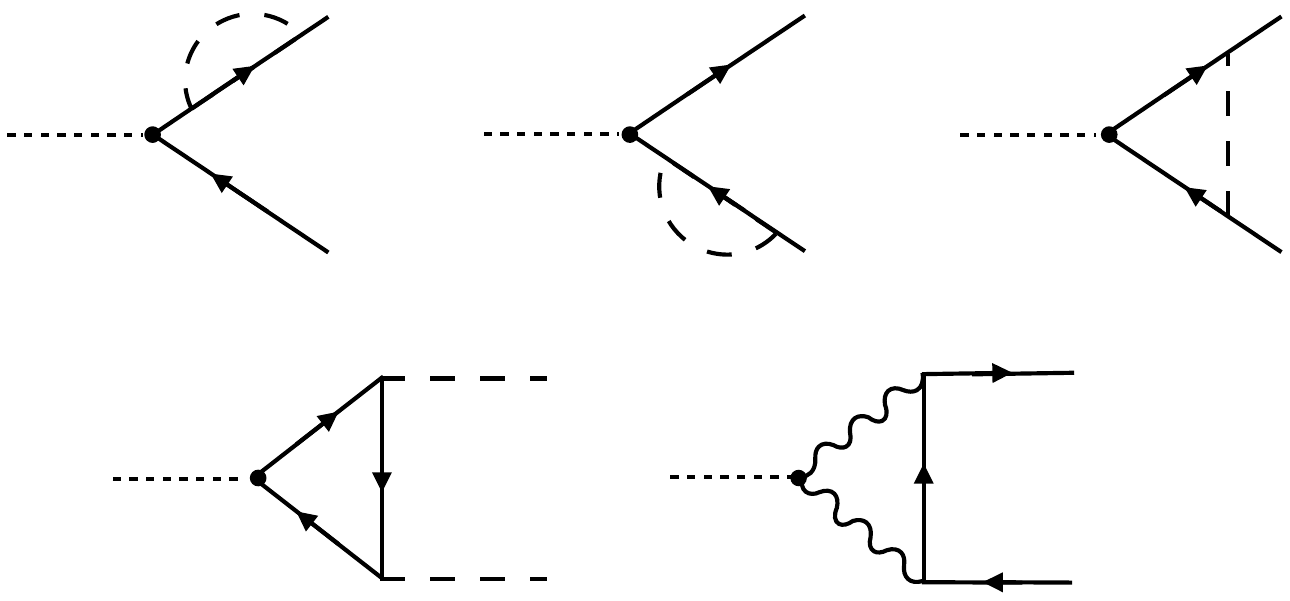}
\caption{\label{fig:mixing} 
One-loop diagrams accounting for operator mixing through Yukawa interactions and gauge interactions.} 
\vspace{-2mm}
\end{center}
\end{figure}

The Wilson coefficients $\bm{c}_F$ of the ALP interactions with fermions in (\ref{Leff}) are scale-dependent quantities and satisfy rather complicated RG equations. At one-loop order there are contributions from Yukawa interactions, which result from the first three graphs shown in Figure~\ref{fig:mixing}. While the external-leg corrections (first two graphs) give rise to multiplicative renormalization effects, which in general are not diagonal in generation space, the vertex diagram (third graph) leads to a mixing of the $SU(2)_L$ singlet and doublet coefficients $\bm{c}_{Q}$ and $\bm{c}_{u,d}$, as well as $\bm{c}_L$ and $\bm{c}_e$. Our results for these contributions to the RG equations agree with the corresponding expressions derived in \cite{Choi:2017gpf,MartinCamalich:2020dfe,Heiles:2020plj}. The first diagram in the second row of Figure~\ref{fig:mixing} shows a class of UV-divergent one-loop diagrams which require the operator $O_\phi$ in (\ref{Higgscoupling}) as a counterterm. As we have discussed in Section~\ref{sec:2} this operator is redundant. It is therefore required to map it back onto our operator basis using the replacement rule (\ref{Ophireplace}). This gives rise to universal contributions in the RG equations proportional to the parameters $\beta_F$ in (\ref{betacond}). In previous studies the operator $O_\phi$ was included as a basis operator, and its coefficient $C_\phi$ not only entered the evolution equations for the ALP--fermion couplings, but in fact was assumed to obey an independent RG equation itself \cite{Choi:2017gpf,MartinCamalich:2020dfe}. Such a treatment gives rise to ambiguous results (see e.g.\ the discussion in Section~3 of \cite{Jenkins:2013zja}), because it is impossible to distinguish the matrix elements of $O_\phi$ from the matrix elements of the fermionic operators $O_F$ in (\ref{Ophireplace}).\footnote{This distinction {\em is\/} possible in related models, in which the analogue of the operator $O_\phi$ is not redundant. An example is provided by the $Z'$ model studied in \cite{Heiles:2020plj}, in which $\partial^\mu a$ in (\ref{Leff}) and (\ref{Higgscoupling}) is replaced by $Z^{\prime\spac\mu}$.} 

In addition, there is a mixing of the Wilson coefficients $c_{VV}$ of the ALP--boson interactions into the coefficients $\bm{c}_F$, shown by the last diagram in Figure~\ref{fig:mixing}. For the case of QCD this mixing has been studied in \cite{Altarelli:1988nr,Chetyrkin:1998mw},\footnote{Note that these authors define the dual field-strength tensor as well as the Levi--Civita symbol differently from us. As a result, their quantity $\tilde G^{\mu\nu,a}$ differs from ours by a factor $(-2)$.} 
and we agree with the findings of these authors. Note that, owing to our normalization of the coefficients $c_{VV}$, the corresponding terms in the evolution equations are proportional to $\alpha_i^2$, and they are diagonal in generation space. Finally, at two-loop order in gauge interactions there are additional generation-independent contributions to the evolution equations, which are proportional to the ALP--fermion couplings. They arise from the second diagram shown in Figure~\ref{fig:2loop} and are diagonal in generation space. We have derived these contributions by generalizing the corresponding results obtained for QCD in \cite{Kodaira:1979pa,Larin:1993tq} to the gauge group of the SM. Combining all effects, we obtain (with $q=u,d$) 
\begin{align}\label{RGEs}
   \frac{d}{d\ln\mu}\,\bm{c}_Q(\mu)
   &= \frac{1}{32\pi^2}\,\big\{ \bm{Y}_u\spac\bm{Y}_u^\dagger 
    + \bm{Y}_d\spac\bm{Y}_d^\dagger, \bm{c}_Q \big\}
    - \frac{1}{16\pi^2}\,\big( \bm{Y}_u\,\bm{c}_u \bm{Y}_u^\dagger 
    + \bm{Y}_d\,\bm{c}_d \bm{Y}_d^\dagger \big) \notag\\
   &\quad\mbox{}+ \left[ \frac{\beta_Q}{8\pi^2}\,X
    - \frac{3\alpha_s^2}{4\pi^2}\,C_F^{(3)}\spac\tilde c_{GG} 
    - \frac{3\alpha_2^2}{4\pi^2}\,C_F^{(2)}\spac\tilde c_{WW}
    - \frac{3\alpha_1^2}{4\pi^2}\,{\cal Y}_Q^2\,\tilde c_{BB} \right] \mathbbm{1} \,, \notag\\
   \frac{d}{d\ln\mu}\,\bm{c}_q(\mu)
   &= \frac{1}{16\pi^2}\,\big\{ \bm{Y}_q^\dagger\spac\bm{Y}_q, \bm{c}_q \big\}
    - \frac{1}{8\pi^2}\,\bm{Y}_q^\dagger\bm{c}_Q \bm{Y}_q 
    + \left[ \frac{\beta_q}{8\pi^2}\,X
    + \frac{3\alpha_s^2}{4\pi^2}\,C_F^{(3)}\spac\tilde c_{GG} 
    + \frac{3\alpha_1^2}{4\pi^2}\,{\cal Y}_q^2\,\tilde c_{BB} \right] \mathbbm{1} \,, \notag\\
   \frac{d}{d\ln\mu}\,\bm{c}_L(\mu)
   &= \frac{1}{32\pi^2} \left\{ \bm{Y}_e\spac\bm{Y}_e^\dagger, \bm{c}_L \right\}
    - \frac{1}{16\pi^2}\,\bm{Y}_e\,\bm{c}_e \bm{Y}_e^\dagger 
    + \left[ \frac{\beta_L}{8\pi^2}\,X
    - \frac{3\alpha_2^2}{4\pi^2}\,C_F^{(2)}\spac\tilde c_{WW} 
    - \frac{3\alpha_1^2}{4\pi^2}\,{\cal Y}_L^2\,\tilde c_{BB} \right] \mathbbm{1} \spac , \notag\\
   \frac{d}{d\ln\mu}\,\bm{c}_e(\mu)
   &= \frac{1}{16\pi^2} \left\{ \bm{Y}_e^\dagger\spac\bm{Y}_e, \bm{c}_e \right\}
    - \frac{1}{8\pi^2}\,\bm{Y}_e^\dagger\bm{c}_L \bm{Y}_e
    + \left[ \frac{\beta_e}{8\pi^2}\,X
    + \frac{3\alpha_1^2}{4\pi^2}\,{\cal Y}_e^2\,\tilde c_{BB} \right] \mathbbm{1} \,,
\end{align}
where $C_F^{(N)}=\frac{N^2-1}{2N}$ is the eigenvalue of the quadratic Casimir operator in the fundamental representation of $SU(N)$, and we have abbreviated 
\begin{equation}\label{Xdef}
   X = \text{Tr} \left[ 
    3\spac\bm{c}_Q\spac\big( \bm{Y}_u\spac\bm{Y}_u^\dagger - \bm{Y}_d\spac\bm{Y}_d^\dagger \big)
    - 3\spac\bm{c}_u\spac\bm{Y}_u^\dagger\spac\bm{Y}_u + 3\spac\bm{c}_d\spac\bm{Y}_d^\dagger\spac\bm{Y}_d
    - \bm{c}_L\spac\bm{Y}_e\spac\bm{Y}_e^\dagger + \bm{c}_e\spac\bm{Y}_e^\dagger\spac\bm{Y}_e \right] .
\end{equation}
All quantities on the right-hand side of (\ref{RGEs}) must be evaluated at the scale $\mu$. Note that the ALP--boson and ALP--fermion couplings entering at ${\cal O}(\alpha_i^2)$ appear precisely in the linear combinations already encountered in (\ref{rela2}), i.e.\
\begin{equation}\label{eq:20}
\begin{aligned}
   \tilde c_{GG} &= c_{GG} + \frac12\,\text{Tr} \left( \bm{c}_u + \bm{c}_d - 2\bm{c}_Q \right) , \\
   \tilde c_{WW} &= c_{WW} - \frac12\,\text{Tr} \left( 3\bm{c}_Q + \bm{c}_L \right) , \\[-0.5mm]
   \tilde c_{BB} &= c_{BB} + \text{Tr} \left( \frac43\,\bm{c}_u + \frac13\,\bm{c}_d
    - \frac16\,\bm{c}_Q + \bm{c}_e - \frac12\,\bm{c}_L \right) .
\end{aligned}
\end{equation}
To the best of our knowledge, the contributions proportional to the quantity $X$, which descend from the redundant operator $O_\phi$, as well as the two-loop contributions to the RG evolution equations for the ALP couplings have been derived here for the first time. The appearance of the coefficients $\beta_F$ in the above relations, which are constrained by the conditions (\ref{betacond}) but are otherwise arbitrary, appears puzzling at first sight. However, all contributions proportional to the unit matrix in the RG equations give rise to flavor-diagonal contributions after transformation to the mass basis. We will see in Sections~\ref{sec:4} and \ref{sec:match} that in predictions for physical quantity any ambiguity in the choice of the $\beta_F$ parameters cancels out.

The relations in (\ref{cVVRGE})\spac--\spac(\ref{eq:20}) form a set of coupled differential equations, from which the scale dependence of the various ALP couplings can be derived. We can simplify the structure of the evolution equations by making use of the freedom to redefine the fermion fields in the SM Lagrangian. The SM Yukawa matrices can be diagonalized by means of bi-unitary transformations, such that 
\begin{equation}\label{Ydiag}
\begin{aligned}
   \bm{U}_u^\dagger\,\bm{Y}_u\,\bm{W_u} 
   &= \bm{Y}_u^{\rm diag} = \text{diag}(y_u,y_c,y_t) \,, \\
   \bm{U}_d^\dagger\,\bm{Y}_d\,\bm{W_d} 
   &= \bm{Y}_d^{\rm diag} = \text{diag}(y_d,y_s,y_b) \,, \\
   \bm{U}_e^\dagger\,\bm{Y}_e\,\bm{W_e} 
   &= \bm{Y}_e^{\rm diag} = \text{diag}(y_e,y_\mu,y_\tau) \,.
\end{aligned}
\end{equation}
If we redefine the fermion fields via
\begin{equation}
\begin{aligned}
   & Q \to \bm{U}_u\,Q \,, \qquad
   u_R \to \bm{W}_u\,u_R \,, \qquad
   d_R \to \bm{W}_d\,d_R \,, \\
   &\hspace{1.75cm} L \to \bm{U}_e\,L \,, \qquad
   e_R \to \bm{W}_e\,e_R \,,
\end{aligned}
\end{equation}
then the up-sector and lepton-sector Yukawa matrices are diagonalized, while the down-sector Yukawa matrix is transformed into
\begin{equation}
   \bm{Y}_d \to \bm{U}_u^\dagger\,\bm{Y}_d\,\bm{W}_d = \bm{V}\,\bm{Y}_d^{\rm diag} \,.
\end{equation}
Here $\bm{V}=\bm{U}_u^\dagger\spac\bm{U}_d$ is the CKM matrix. For the purposes of the following discussion we define the matrices $\bm{c}_F$ in this particular basis of fields. Moreover, because of the smallness of the masses of the SM fermions except the top quark, it is a very good approximation to neglect all Yukawa couplings other than $y_t\simeq 1$.\footnote{Since the different Yukawa matrices appear in pairs in (\ref{RGEs}), the contributions of the Yukawa couplings of the bottom quark or the $\tau$ lepton would be suppressed, relative to the $y_t^2$ terms, by factors of $y_b^2/y_t^2\sim y_\tau^2/y_t^2\sim 10^{-4}$. Some of the two-loop electroweak contributions included in the RG equations are of a similar magnitude; however, as explained earlier, we keep these effects because they are proportional to the ALP--boson couplings and hence can be enhanced in some ALP models.} 
The RG equations for the ALP--fermion couplings then simplify to 
\begin{equation}\label{RGEsbasis}
\begin{aligned}
   \frac{d}{d\ln\mu}\,\big[ c_Q(\mu) \big]_{ii}
   &= - \frac{y_t^2}{8\pi^2} \left( \frac{\delta_{i3}}{2} + 3\beta_Q \right) c_{tt}
    - \frac{\alpha_s^2}{\pi^2}\,\tilde c_{GG} - \frac{9\alpha_2^2}{16\pi^2}\,\tilde c_{WW} 
    - \frac{\alpha_1^2}{48\pi^2}\,\tilde c_{BB} \,, \\
   \frac{d}{d\ln\mu}\,\big[ c_Q(\mu) \big]_{ij}
   &= \frac{y_t^2}{32\pi^2} \left( \delta_{i3} + \delta_{j3} \right) \left( c_Q \right)_{ij} ;
    \quad i\ne j \,, \\
   \frac{d}{d\ln\mu}\,\big[ c_u(\mu) \big]_{ii}
   &= \frac{y_t^2}{8\pi^2} \left( \delta_{i3} - 3\beta_u \right) c_{tt}
    + \frac{\alpha_s^2}{\pi^2}\,\tilde c_{GG} 
    + \frac{\alpha_1^2}{3\pi^2}\,\tilde c_{BB} \,, \\
   \frac{d}{d\ln\mu}\,\big[ c_u(\mu) \big]_{ij}
   &= \frac{y_t^2}{16\pi^2} \left( \delta_{i3} + \delta_{j3} \right) \left( c_u \right)_{ij} ; 
    \quad i\ne j \,, \\
   \frac{d}{d\ln\mu}\,\big[ c_d(\mu) \big]_{ij}
   &= \delta_{ij} \left( - \frac{3 y_t^2}{8\pi^2}\,\beta_d\,c_{tt}
    + \frac{\alpha_s^2}{\pi^2}\,\tilde c_{GG}
    + \frac{\alpha_1^2}{12\pi^2}\,\tilde c_{BB} \right) , \\
   \frac{d}{d\ln\mu}\,\big[ c_L(\mu) \big]_{ij}
   &= \delta_{ij} \left( - \frac{3 y_t^2}{8\pi^2}\,\beta_L\,c_{tt} 
    - \frac{9\alpha_2^2}{16\pi^2}\,\tilde c_{WW} 
    - \frac{3\alpha_1^2}{16\pi^2}\,\tilde c_{BB} \right) , \\
   \frac{d}{d\ln\mu}\,\big[ c_e(\mu) \big]_{ij}
   &= \delta_{ij} \left( - \frac{3 y_t^2}{8\pi^2}\,\beta_e\,c_{tt}
    + \frac{3\alpha_1^2}{4\pi^2}\,\tilde c_{BB} \right) .
\end{aligned}
\end{equation}
where we have defined
\begin{equation}\label{cttdef}
   c_{tt}(\mu) = \left[c_u(\mu)\right]_{33} - \left[c_Q(\mu)\right]_{33} .
\end{equation}
With our choice of the basis of fermion fields, this quantity will turn out to be the coupling of the ALP to the physical top-quark mass eigenstate (see Section~\ref{sec:4} below).

\subsection{General solution of the evolution equations}
\label{subsec:RGEsolutions}

Whereas the original ALP--boson couplings $c_{VV}$ are scale independent, this is no longer true for the couplings $\tilde c_{VV}$, whose definitions contain the scale-dependent ALP--fermion couplings. This fact is discussed in more detail in Appendix~\ref{app:A}. We find that (in the approximation where only the top-quark Yukawa coupling is kept, see above) the four functions $\tilde c_{GG}(\mu)$, $\tilde c_{WW}(\mu)$, $\tilde c_{BB}(\mu)$ and $c_{tt}(\mu)$ satisfy a closed set of coupled differential equations, which can be solved. We obtain
\begin{equation}\label{cVVevol}
\begin{aligned}
   \tilde c_{GG}(\mu) 
   &= \tilde c_{GG}(\Lambda) - \frac29 \left( 1 - e^{-18\spac U(\mu,\Lambda)} \right) c_{tt}(\Lambda) \,, \\
   \tilde c_{WW}(\mu) 
   &= \tilde c_{WW}(\Lambda) - \frac16 \left( 1 - e^{-18\spac U(\mu,\Lambda)} \right) c_{tt}(\Lambda) \,, \\
   \tilde c_{BB}(\mu) 
   &= \tilde c_{BB}(\Lambda) - \frac{17}{54} \left( 1 - e^{-18\spac U(\mu,\Lambda)} \right) 
    c_{tt}(\Lambda) \,,
\end{aligned}
\end{equation}
and
\begin{equation}\label{cQcu2}
\begin{aligned}
   c_{tt}(\mu) 
   &= e^{-18\spac U(\mu,\Lambda)}\,c_{tt}(\Lambda) \\
   &\quad\hbox{}+ \int_\Lambda^{\mu}\!\frac{d\mu'}{\mu'}\,e^{-18\spac U(\mu,\mu')}
    \left[\spac \frac{2\alpha_s^2(\mu')}{\pi^2}\,\tilde c_{GG}(\mu')
    + \frac{9\alpha_2^2(\mu')}{16\pi^2}\,\tilde c_{WW}(\mu')
    + \frac{17\alpha_1^2(\mu')}{48\pi^2}\,\tilde c_{BB}(\mu') \right] .
\end{aligned}
\end{equation}
where 
\begin{equation}\label{Udef}
   U(\mu,\Lambda)
   = - \int_{\Lambda}^{\mu}\!\frac{d\mu'}{\mu'}\,\frac{y_t^2(\mu')}{32\pi^2}
\end{equation}
is defined in terms of the running top-quark Yukawa coupling. In the solutions (\ref{cVVevol}) we neglect higher-order terms such as those shown in the second line of (\ref{cQcu2}), which is consistent because the effective ALP--boson couplings enter only at two-loop order in (\ref{RGEs}).

Using these solutions, we can now integrate the equations (\ref{RGEsbasis}) to obtain the evolution of the various ALP--fermion couplings from the new-physics scale $\Lambda$ down to the weak-interaction scale $\mu_w\sim 100$\,GeV. For example, the solution of the first equation takes the form
\begin{equation}
\begin{aligned}
   \left[ c_Q(\mu_w) \right]_{ii} 
   &= \left[ c_Q(\Lambda) \right]_{ii} 
    - \left( \beta_Q + \frac{\delta_{i3}}{6} \right) I_t(\mu_w,\Lambda) \\
   &\quad\mbox{}- \int_\Lambda^{\mu_w}\!\frac{d\mu}{\mu}
    \left[ \frac{\alpha_s^2(\mu)}{\pi^2}\,\tilde c_{GG}(\mu)
    + \frac{9\alpha_2^2(\mu)}{16\pi^2}\,\tilde c_{WW}(\mu) 
    + \frac{\alpha_1^2(\mu)}{48\pi^2}\,\tilde c_{BB}(\mu) \right] ,
\end{aligned}
\end{equation}
where we have defined, using relation (\ref{cQcu2}),
\begin{equation}\label{Itdef}
\begin{aligned}
   I_t(\mu_w,\Lambda) 
   &\equiv \int_\Lambda^{\mu_w}\!\frac{d\mu}{\mu}\,\frac{3 y_t^2(\mu)}{8\pi^2}\,c_{tt}(\mu) \\
   &= - \frac23 \left( 1 - e^{-18\spac U(\mu_w,\Lambda)} \right) c_{tt}(\Lambda) \\
   &\quad\mbox{}- \int_\Lambda^{\mu_w}\!\frac{d\mu}{\mu}
    \left( 1 - e^{-18\spac U(\mu_w,\mu)} \right)\!
    \left[ \frac{4\alpha_s^2(\mu)}{3\pi^2}\,\tilde c_{GG}(\mu)
    + \frac{3\alpha_2^2(\mu)}{8\pi^2}\,\tilde c_{WW}(\mu)
    + \frac{17\alpha_1^2(\mu)}{72\pi^2}\,\tilde c_{BB}(\mu) \right] \spac\! .
\end{aligned}
\end{equation}
In this way all results can be expressed in terms of $U(\mu,\Lambda)$ and integrals over the running gauge couplings with the ALP--boson couplings $\tilde c_{VV}(\mu)$ in (\ref{cVVevol}). The scale evolution of the gauge couplings is governed by the set of coupled differential equations $d\alpha_i(\mu)/d\ln\mu=\beta^{(i)}(\{\alpha_j\})$, where the $\beta$-functions of the three gauge groups are of the form
\begin{equation}
   \beta^{(i)}(\{\alpha_j\}) 
   = - \beta_0^{(i)}\,\frac{\alpha_i^2}{2\pi} + {\cal O}(\alpha_i^2\alpha_j) \,.
\end{equation}
Above the weak scale the relevant one-loop coefficients are $\beta_0^{(1)}=-\frac{41}{6}$, $\beta_0^{(2)}=\frac{19}{6}$ and $\beta_0^{(3)}=7$. Starting at two-loop order mixed terms appear, where $\alpha_j\in\{\alpha_1,\alpha_2,\alpha_s,\alpha_t,\frac{\lambda}{4\pi}\}$ with $\alpha_t=y_t^2/(4\pi)$ can be any one of the SM coupling parameters. The complete three-loop expressions for the $\beta$-functions can be found in \cite{Mihaila:2012fm}.\footnote{The coupling parameter $\alpha_1$ in this work differs from our $\alpha_1$ by a factor 5/3.} 

We now present our final expressions for the RG-evolved ALP--fermion couplings, beginning with flavor non-diagonal effects, which are insensitive to the $\beta_F$ parameters. We find (with $i\ne j$)
\begin{equation}\label{cijsolu}
\begin{aligned}
   \left[ c_Q(\mu_w) \right]_{ij}
   &= e^{-(\delta_{i3} + \delta_{j3})\,U(\mu_w,\Lambda)} \left[ c_Q(\Lambda) \right]_{ij} , \\
   \left[ c_u(\mu_w) \right]_{ij}
   &= e^{-2(\delta_{i3} + \delta_{j3})\,U(\mu_w,\Lambda)} \left[ c_u(\Lambda) \right]_{ij} , \\
   \left[ c_d(\mu_w) \right]_{ij} 
   &= \left[ c_d(\Lambda) \right]_{ij} , \\
   \left[ c_L(\mu_w) \right]_{ij} 
   &= \left[ c_L(\Lambda) \right]_{ij} , \\
   \left[ c_e(\mu_w) \right]_{ij} 
   &= \left[ c_e(\Lambda) \right]_{ij} ,
\end{aligned}
\end{equation}
as well as
\begin{equation}\label{cQ33cQ11}
\begin{aligned}
   \left[ c_Q(\mu_w) \right]_{33} - \left[ c_Q(\mu_w) \right]_{11}
   &= \left[ c_Q(\Lambda) \right]_{33} - \left[ c_Q(\Lambda) \right]_{11} 
    - \frac16\,I_t(\mu_w,\Lambda) \,, \\
   \left[ c_u(\mu_w) \right]_{33} - \left[ c_u(\mu_w) \right]_{11}
   &= \left[ c_u(\Lambda) \right]_{33} - \left[ c_u(\Lambda) \right]_{11} 
    + \frac13\,I_t(\mu_w,\Lambda) \,.
\end{aligned}
\end{equation}
The last two relations show how a possible flavor non-universality of the diagonal couplings $\left[c_{Q,u}(\Lambda)\right]_{ii}$ at the new-physics scale, which is allowed even under the assumption of minimal flavor violation, evolves to low energies.

Before presenting our solutions for the generation-diagonal couplings we return to the question of the $\beta_F$ dependence of the evolution equations (\ref{RGEsbasis}), which hints at a redundancy of our results. In all physical quantities the dependence on these parameters cancels out. It follows that only certain linear combinations of the flavor-diagonal ALP--fermion couplings are physical. In particular, we find that the differences
\begin{align}\label{cffsolu}
   \left[ c_u(\mu_w) \right]_{ii} - \left[ c_Q(\mu_w) \right]_{ii}
   &= \left[ c_u(\Lambda) \right]_{ii} - \left[ c_Q(\Lambda) \right]_{ii} 
    + \left( 1 + \frac{\delta_{i3}}{2} \right) I_t(\mu_w,\Lambda) \notag\\
   &\quad\hbox{}+ \int_\Lambda^{\mu_w}\!\frac{d\mu}{\mu}
    \left[ \frac{2\alpha_s^2(\mu)}{\pi^2}\,\tilde c_{GG}(\mu)
    + \frac{9\alpha_2^2(\mu)}{16\pi^2}\,\tilde c_{WW}(\mu)
    + \frac{17\alpha_1^2(\mu)}{48\pi^2}\,\tilde c_{BB}(\mu) \right] , \notag
\end{align}
% forced page break
\begin{align}
   \left[ c_d(\mu_w) \right]_{ii} - \left[ c_Q(\mu_w) \right]_{ii}
   &= \left[ c_d(\Lambda) \right]_{ii} - \left[ c_Q(\Lambda) \right]_{ii} 
    - \left( 1 - \frac{\delta_{i3}}{6} \right) I_t(\mu_w,\Lambda) \notag\\
   &\quad\hbox{}+ \int_\Lambda^{\mu_w}\!\frac{d\mu}{\mu}
    \left[ \frac{2\alpha_s^2(\mu)}{\pi^2}\,\tilde c_{GG}(\mu)
    + \frac{9\alpha_2^2(\mu)}{16\pi^2}\,\tilde c_{WW}(\mu)
    + \frac{5\alpha_1^2(\mu)}{48\pi^2}\,\tilde c_{BB}(\mu) \right] , \notag\\
   \left[ c_e(\mu_w) \right]_{ii} - \left[ c_L(\mu_w) \right]_{ii}
   &= \left[ c_e(\Lambda) \right]_{ii} - \left[ c_L(\Lambda) \right]_{ii} - I_t(\mu_w,\Lambda) \notag\\
   &\quad\hbox{}+ \int_\Lambda^{\mu_w}\!\frac{d\mu}{\mu}
    \left[ \frac{9\alpha_2^2(\mu)}{16\pi^2}\,\tilde c_{WW}(\mu)
    + \frac{15\alpha_1^2(\mu)}{16\pi^2}\,\tilde c_{BB}(\mu) \right] 
\end{align}
are independent of the $\beta_F$ parameters once the relations (\ref{betacond}) are taken into account. For $i=3$ the first relation reduces to (\ref{cQcu2}). We will see in Section~\ref{sec:4} that the solutions (\ref{cijsolu})\spac--\spac(\ref{cffsolu}) are sufficient to calculate arbitrary physical processes involving ALPs, where however the second relation in (\ref{cffsolu}) gets modified when one transforms the left-handed down-quark fields to the mass basis.

We can push further and obtain an explicit approximate expression for the quantity $U(\mu_w,\Lambda)$ in (\ref{Udef}). At leading order in perturbation theory the top-quark Yukawa coupling and the strong coupling obey the coupled system of equations 
\begin{equation}
   \frac{d\alpha_s(\mu)}{d\ln\mu} = - \frac{7\alpha_s^2(\mu)}{2\pi} \,, \qquad
   \frac{d\alpha_t(\mu)}{d\ln\mu} = \frac{\alpha_t(\mu)}{2\pi}
    \left[ \frac92\,\alpha_t(\mu) - 8\alpha_s(\mu) \right] ,  
\end{equation}
where we neglect the small effects of the weak interactions. The exact solution of this system exhibits a ``quasi fixed point'', where the running of $\alpha_t(\mu)$ tracks the evolution of $\alpha_s(\mu)$. One finds \cite{Pendleton:1980as,Hill:1980sq}
\begin{equation}
   \frac{\alpha_t(\mu)}{\alpha_t(\mu_0)}
   = \left( \frac{\alpha_s(\mu)}{\alpha_s(\mu_0)} \right)^{\!\frac87}
    \left[ 1 + \frac92\,\frac{\alpha_t(\mu_0)}{\alpha_s(\mu_0)}\,
    \bigg[ \left( \frac{\alpha_s(\mu)}{\alpha_s(\mu_0)} \right)^{\!\frac17} - 1 \bigg]
    \right]^{-1} \!,
\end{equation}
where $\mu_0$ is some reference scale. Using this result in (\ref{Udef}), we find after a straightforward calculation
\begin{equation}\label{Usolu}
   U(\mu,\Lambda)
   = - \frac{1}{18}\,\ln\!\Bigg[ 1 - \frac92\,\frac{\alpha_t(\mu)}{\alpha_s(\mu)}\,
    \bigg[ 1 - \left( \frac{\alpha_s(\Lambda)}{\alpha_s(\mu)} \right)^{\!\frac17} \bigg] \Bigg] 
   \approx \frac{y_t^2(\mu)}{64\pi^2}\,\ln\frac{\Lambda^2}{\mu^2} + \dots \,,
\end{equation}
where the dots refer to terms of order $y_t^4\ln^2(\Lambda^2/\mu^2)$ and higher. In this expression the large logarithms of the scale ratio $\Lambda^2/\mu^2$ are resummed to all orders of perturbation theory. This explicit result implies that
\begin{equation}
   1 - e^{-18\spac U(\mu_w,\Lambda)}
   = \frac92\,\frac{\alpha_t(\mu_w)}{\alpha_s(\mu_w)}\,
    \bigg[ 1 - \left( \frac{\alpha_s(\Lambda)}{\alpha_s(\mu_w)} \right)^{\!\frac17} \bigg] \,.
\end{equation}
The numerical impact of the evolution effects on the ALP--fermion couplings will be discussed in Section~\ref{sec:match}.

\section{Transformation to the mass basis}
\label{sec:4}

Once the effective Lagrangian has been evolved to the weak scale $\mu_w$, it is appropriate to express it in terms of fields defined in the broken phase of the electroweak symmetry, which correspond to the mass eigenstates of physical particles. This leads to
\begin{equation}\label{LeffmuW}
\begin{aligned}
   {\cal L}_{\rm eff}(\mu_w)
   &= \frac12 \left( \partial_\mu a\right)\!\left( \partial^\mu a\right) - \frac{m_{a,0}^2}{2}\,a^2
    + {\cal L}_{\rm ferm}(\mu_w) 
    + c_{GG}\,\frac{\alpha_s}{4\pi}\,\frac{a}{f}\,G_{\mu\nu}^a\,\tilde G^{\mu\nu,a}
    + c_{\gamma\gamma}\,\frac{\alpha}{4\pi}\,\frac{a}{f}\,F_{\mu\nu}\,\tilde F^{\mu\nu} \\
   &\quad\mbox{}+ c_{\gamma Z}\,\frac{\alpha}{2\pi s_w\spac c_w}\,
    \frac{a}{f}\,F_{\mu\nu}\,\tilde Z^{\mu\nu}
    + c_{ZZ}\,\frac{\alpha}{4\pi s_w^2\spac c_w^2}\,\frac{a}{f}\,Z_{\mu\nu}\,\tilde Z^{\mu\nu} 
    + c_{WW}\,\frac{\alpha}{2\pi s_w^2}\,\frac{a}{f}\,W_{\mu\nu}^+\,\tilde W^{-\mu\nu} \,,
\end{aligned}
\end{equation}
where $s_w\equiv\sin\theta_W$ and $c_w\equiv\cos\theta_W$ denote the sine and cosine of the weak mixing angle, and we have defined \cite{Bauer:2017ris}
\begin{equation}\label{eq:30}
   c_{\gamma\gamma} = c_{WW} + c_{BB} \,, \qquad
   c_{\gamma Z} = c_w^2\,c_{WW} - s_w^2\,c_{BB} \,, \qquad
   c_{ZZ} = c_w^4\,c_{WW} + s_w^4\,c_{BB} \,.
\end{equation}
All coupling parameters and operators in (\ref{LeffmuW}) are now defined at the weak scale $\mu_w$. Recall that the Wilson coefficients $c_{VV}$ are scale independent.

To obtain the ALP interactions with fermions contained in ${\cal L}_{\rm ferm}$ we must transform the fermion fields to the mass basis, in which the Yukawa matrices are diagonalized, see (\ref{Ydiag}). Under the corresponding field redefinitions the flavor matrices $\bm{c}_F$ transform into new hermitian matrices
\begin{equation}\label{kFkfdef}
\begin{aligned}
   \bm{k}_U &= \bm{U}_u^\dagger\spac\bm{c}_Q\spac\bm{U}_u \,, \qquad
    \bm{k}_D = \bm{U}_d^\dagger\,\bm{c}_Q\spac\bm{U}_d \,, \qquad
    \bm{k}_E = \bm{U}_e^\dagger\spac\bm{c}_L\spac\bm{U}_e \,, \\
   &\hspace{2.0cm} \bm{k}_f = \bm{W}_f^\dagger\spac\bm{c}_f \bm{W}_f \,; \quad f=u,d,e \,.
\end{aligned}
\end{equation}
Note that the two matrices $\bm{k}_U$ and $\bm{k}_D$ are connected via the CKM matrix $\bm{V}$, such that 
\begin{equation}
   \bm{k}_D = \bm{V}^\dagger\bm{k}_U \bm{V} \spac ,
\end{equation}
and are therefore not independent. Likewise, the ALP couplings to the neutrinos are identical to those to the left-handed charged leptons, i.e.\ $\bm{k}_\nu=\bm{k}_E$. In terms of these matrices we obtain
\begin{equation}\label{Lferm}
\begin{aligned}
   {\cal L}_{\rm ferm}(\mu_w)
   &= \frac{\partial^\mu a}{f}\,\Big[
    \bar u_L\spac\bm{k}_U\spac\gamma_\mu\spac u_L + \bar u_R\spac\bm{k}_u\spac\gamma_\mu\spac u_R 
    + \bar d_L\spac\bm{k}_D\spac\gamma_\mu\spac d_L + \bar d_R\spac\bm{k}_d\spac\gamma_\mu\spac d_R \\
   &\hspace{1.45cm}\mbox{}+ \bar\nu_L\spac\bm{k}_\nu\spac\gamma_\mu\spac\nu_L 
    + \bar e_L\spac\bm{k}_E\spac\gamma_\mu\spac e_L + \bar e_R\spac\bm{k}_e\spac\gamma_\mu\spac e_R \Big] \,.
\end{aligned}
\end{equation}
The matrices $\bm{k}_F$ and $\bm{k}_f$ are evaluated at the scale $\mu_w$. The corresponding expressions can be obtained from the results compiled in Section~\ref{subsec:RGEsolutions} by recalling that these relations have been derived in a basis for which all transformation matrices are equal to the unit matrix except for $\bm{U}_d=\bm{V}$. It thus follows that $\bm{k}_U=\bm{c}_Q$, $\bm{k}_E=\bm{k}_\nu=\bm{c}_L$, $\bm{k}_{u,d,e}=\bm{c}_{u,d,e}$, while $\bm{k}_D=\bm{V}^\dagger\bm{c}_Q \bm{V}$. 

It is instructive to study what the hypothesis of minimal flavor violation \cite{DAmbrosio:2002vsn} implies for the structure of the ALP--fermion couplings after electroweak symmetry breaking. Transforming the expressions (\ref{MFV1}) to the mass basis, we obtain
\begin{equation}
\begin{aligned}
   \bm{k}_U &= c_0^Q\,\mathbbm{1} + \epsilon \left[ c_1^Q\,(\bm{Y}_u^{\rm diag})^2 
    + c_2^Q\,\bm{V}\spac(\bm{Y}_d^{\rm diag})^2\spac\bm{V}^\dagger \right] + {\cal O}(\epsilon^2) \,, \\
   \bm{k}_D &= c_0^Q\,\mathbbm{1} + \epsilon \left[ 
    c_1^Q\,\bm{V}^\dagger\spac(\bm{Y}_u^{\rm diag})^2\spac\bm{V} + c_2^Q\,(\bm{Y}_d^{\rm diag})^2 \right]
    + {\cal O}(\epsilon^2) \,, \\
   \bm{k}_u &= c_0^u\,\mathbbm{1} + \epsilon\,c_1^u\,(\bm{Y}_u^{\rm diag})^2
    + \epsilon^2 \left[ c_2^u\,(\bm{Y}_u^{\rm diag})^4
    + c_3^u\,\bm{Y}_u^{\rm diag}\,\bm{V}\spac(\bm{Y}_d^{\rm diag})^2\spac\bm{V}^\dagger\,
     \bm{Y}_u^{\rm diag} \right] + {\cal O}(\epsilon^3) \,, \\
   \bm{k}_d &= c_0^d\,\mathbbm{1} + \epsilon\,c_1^d\,(\bm{Y}_d^{\rm diag})^2
    + \epsilon^2 \left[ c_2^d\,(\bm{Y}_d^{\rm diag})^4
    + c_3^d\,\bm{Y}_d^{\rm diag}\,\bm{V}^\dagger\spac(\bm{Y}_u^{\rm diag})^2\spac\bm{V}\,
     \bm{Y}_d^{\rm diag} \right] + {\cal O}(\epsilon^3) \,.
\end{aligned}
\end{equation}
The only non-diagonal contributions are those involving the CKM matrix. To very good approximation we can set the diagonal entries of the Yukawa matrices to zero for all quarks other than the top quark. In this approximation 
\begin{equation}\label{MFV}
\begin{aligned}
   \bm{k}_U &= c_0^Q\,\mathbbm{1} + \epsilon\,c_1^Q\,(\bm{Y}_t)^2 + {\cal O}(\epsilon^2) \,, \\
   \bm{k}_D &= c_0^Q\,\mathbbm{1} + \epsilon\,c_1^Q\,\bm{V}^\dagger\,(\bm{Y}_t)^2\,\bm{V}
    + {\cal O}(\epsilon^2) \,, \\
   \bm{k}_u &= c_0^u\,\mathbbm{1} + \epsilon\,c_1^u\,(\bm{Y}_t)^2 + {\cal O}(\epsilon^2) \,, \\
   \bm{k}_d &= c_0^d\,\mathbbm{1} \,,
\end{aligned}
\end{equation}
with $\bm{Y}_t=\mbox{diag}(0,0,y_t)$. Note that $[\bm{V}^\dagger\,(\bm{Y}_t)^n\,\bm{V}]_{ij}=y_t^n\,V_{3i}^*\,V_{3j}$. Higher-order terms in $\epsilon$ have the effect of generating more complicated functions of the top-quark mass, while the dependence on CKM parameters remains unchanged. We thus find that, under the hypothesis of minimal flavor violation and to very good approximation, flavor-violating couplings only arise in the couplings $\bm{k}_D$ to left-handed down-type quark currents. The leptonic couplings $\bm{k}_E$ and $\bm{k}_e$ are proportional to the unit matrix in this approximation. 

Several important weak-scale processes involving ALPs have been discussed in the literature \cite{Chen:2010su,Mimasu:2014nea,Jaeckel:2015jla,Knapen:2016moh,Brivio:2017ije,Bauer:2017nlg,Bauer:2017ris,Knapen:2017ebd,Mariotti:2017vtv,Craig:2018kne,Bauer:2018uxu,Aloni:2018vki,Alonso-Alvarez:2018irt,Aloni:2019ruo,Baldenegro:2019whq,Gavela:2019cmq,Coelho:2020saz}. Their rates can be calculated in terms of the couplings entering the effective weak-scale Lagrangian (\ref{LeffmuW}). To mention three prominent examples, we briefly consider the decay $a\to\gamma\gamma$ of a heavy ALP (with mass of order the weak scale) as well as the exotic decay modes $Z\to\gamma a$ and $h\to Za$ of the $Z$ boson and the Higgs boson. Calculating the corresponding decay amplitudes at one-loop order, and setting the matching scale $\mu_w$ equal to the mass of the decaying particle, one finds \cite{Bauer:2017ris,Bauer:2016ydr,Bauer:2016zfj}
\begin{equation}\label{rates}
\begin{aligned}
   \Gamma(a\to\gamma\gamma) 
   &= \frac{\alpha^2\spac m_a^3}{64\pi^3 f^2}\,\big| C_{\gamma\gamma}^{\rm eff} \big|^2 \,, \\
   \Gamma(Z\to\gamma a) 
   &= \frac{m_Z^3}{96\pi^3 f^2}\,
    \frac{\alpha\,\alpha(m_Z)}{s_w^2\spac c_w^2}\,\big| C_{\gamma Z}^{\rm eff} \big|^2
    \left( 1 - \frac{m_a^2}{m_Z^2} \right)^3 , \\
   \Gamma(h\to Za) 
   &= \frac{9\hspace{0.3mm} m_h^3}{256\pi^3 f^2}\,\alpha_t^2(m_h)\,c_{tt}^2(m_h)\,F^2\, 
    \lambda^{3/2}\bigg(\frac{m_Z^2}{m_h^2},\frac{m_a^2}{m_h^2}\bigg) \,.   
\end{aligned}
\end{equation}
The coefficients $C_{\gamma\gamma}^{\rm eff}$ and $C_{\gamma Z}^{\rm eff}$ in the first two cases are given by
\begin{equation}\label{eq:47a}
\begin{aligned}
   C_{\gamma\gamma}^{\rm eff}
   &= c_{\gamma\gamma} + \sum_f N_c^f\spac Q_f^2\,c_{ff}(m_a)\,
    B_1\bigg(\frac{4m_f^2}{m_a^2}\bigg)
    + \frac{2\alpha}{\pi}\,\frac{c_{WW}}{s_w^2}\,B_2\bigg(\frac{4m_W^2}{m_a^2}\bigg) \,, \\
   C_{\gamma Z}^{\rm eff}
   &= c_{\gamma Z} + \sum_f N_c^f\spac Q_f \left( \frac12\,T_3^f-Q_f\spac s_w^2 \right) 
    c_{ff}(m_Z)\spac B_3\bigg(\frac{4m_f^2}{m_a^2},\frac{4m_f^2}{m_Z^2}\bigg) \,,
\end{aligned}
\end{equation}
where $Q_f$ and $N_c^f$ are the electric charges (in units of $e$) and number of colors of the SM fermions (quarks and leptons), $T_3^f$ denotes the weak isospin of the left-handed component of the fermion $f$, and the sum runs over all SM fermion mass eigenstates. The relevant loop functions read 
\begin{equation}\label{B2def}
\begin{aligned}
   B_1(\tau) &= 1 - \tau\,f^2(\tau) \,, \qquad
    B_2(\tau) = 1 - (\tau-1)\,f^2(\tau) \,, \\
   B_3(\tau_1,\tau_2) &= 1 + \frac{\tau_1\spac\tau_2}{\tau_1-\tau_2}
    \left[ f^2(\tau_1) - f^2(\tau_2) \right] ,
\end{aligned}
\end{equation}
with $f(\tau)$ as defined in (\ref{B1def}). 
The function $B_1\approx 1$ for all light fermions with mass $m_f\ll m_a$, while $B_1\approx-\frac{m_a^2}{12m_f^2}$ for heavy fermions ($m_f\gg m_a$). Thus, each electrically charged fermion lighter than the ALP adds a potentially large contribution to the effective Wilson coefficient $C_{\gamma\gamma}^{\rm eff}$, while fermions heavier than the ALP decouple. Similarly, one finds that $B_3\approx 1$ for all fermions much lighter than the $Z$ boson (irrespective of the ALP mass), while for the top quark $|B_3|\ll 1$ as long as the ALP is lighter than the top-quark mass. In the third decay rate in (\ref{rates}) we have defined the phase-space function $\lambda(x,y)=(1-x-y)^2-4xy$ and the parameter integral 
\begin{equation}
   F = \int_0^1\!d[xyz]\,\frac{2m_t^2-x m_h^2-z m_Z^2}{m_t^2-xy m_h^2-yz m_Z^2-xz m_a^2} \,,
\end{equation}
where $d[xyz]\equiv dx\,dy\,dz\,\delta(1-x-y-z)$. Throughout this paper $m_t\equiv\overline{m}_t(m_t)$ denotes the running top-quark mass in the $\overline{\rm MS}$ scheme evaluated at $\mu=m_t$. 
The quantity $F$ is numerically close to 1 for ALP masses below the weak scale. Finally, we have introduced the parameters
\begin{equation}\label{cffdef}
   c_{f_i f_i}(\mu) = \left[ k_f(\mu) \right]_{ii} - \left[ k_F(\mu) \right]_{ii} ,
\end{equation}
which contain the relevant ALP couplings to fermions and will play an important role in our discussion below. This definition generalizes relation (\ref{cttdef}) for the top quark to other ALP--fermion couplings. 

The scale evolution of these quantities from the new-physics scale $\Lambda$ to the electroweak scale can be derived from (\ref{cffsolu}). For up-type quarks and charged leptons, the parameters $c_{ff}$ are equal to the differences of ALP--fermion couplings considered in this result, and we have
\begin{equation}\label{eq:51}
\begin{aligned}
   c_{u_i u_i}(\mu_w) 
   &= c_{u_i u_i}(\Lambda) + \left( 1 + \frac{\delta_{i3}}{2} \right) I_t(\mu_w,\Lambda) \\
   &\quad\hbox{}+ \int_\Lambda^{\mu_w}\!\frac{d\mu}{\mu}
    \left[ \frac{2\alpha_s^2(\mu)}{\pi^2}\,\tilde c_{GG}(\mu)
    + \frac{9\alpha_2^2(\mu)}{16\pi^2}\,\tilde c_{WW}(\mu)
    + \frac{17\alpha_1^2(\mu)}{48\pi^2}\,\tilde c_{BB}(\mu) \right] , \\
   c_{e_i e_i}(\mu_w) 
   &= c_{e_i e_i}(\Lambda) - I_t(\mu_w,\Lambda) 
    + \int_\Lambda^{\mu_w}\!\frac{d\mu}{\mu}
    \left[ \frac{9\alpha_2^2(\mu)}{16\pi^2}\,\tilde c_{WW}(\mu)
    + \frac{15\alpha_1^2(\mu)}{16\pi^2}\,\tilde c_{BB}(\mu) \right] .
\end{aligned}
\end{equation}
For down-type quarks one finds that 
\begin{equation}
   c_{d_i d_i}(\mu_w) 
   = \left[ c_d(\mu_w) \right]_{ii} - \left[ V^\dagger c_Q(\mu_w)\spac V \right]_{ii} ,
\end{equation}
where $\bm{V}$ is the CKM matrix. Hence, the result given in (\ref{cffsolu}) is not directly applicable. Instead, we obtain
\begin{equation}\label{eq:52}
\begin{aligned}
   c_{d_i d_i}(\mu_w) 
   &= c_{d_i d_i}(\Lambda) - I_t(\mu_w,\Lambda) + \frac{|V_{3i}|^2}{6}\,I_t(\mu_w,\Lambda) \\[-1mm]
   &\quad\hbox{}+ \int_\Lambda^{\mu_w}\!\frac{d\mu}{\mu}
    \left[ \frac{2\alpha_s^2(\mu)}{\pi^2}\,\tilde c_{GG}(\mu)
    + \frac{9\alpha_2^2(\mu)}{16\pi^2}\,\tilde c_{WW}(\mu)
    + \frac{5\alpha_1^2(\mu)}{48\pi^2}\,\tilde c_{BB}(\mu) \right] \\[1mm]
   &\quad\mbox{}+ V_{mi}^* V_{ni} \left( \delta_{m3} + \delta_{n3} - 2\spac\delta_{m3}\spac\delta_{n3} \right)
    \left( 1 - e^{-U(\mu_w,\Lambda)} \right) \left[ k_U(\Lambda) \right]_{mn} .
\end{aligned}
\end{equation}
If the matrix $\bm{k}_U(\Lambda)$ is diagonal, as required under the hypothesis of minimal flavor violation, see (\ref{MFV}), then the terms shown in the third line vanish.

\section{Matching contributions at the weak scale}
\label{sec:match}

Let us now assume that the ALP is significantly lighter than the weak scale, and that we are interested in low-energy processes at energies $E\ll 100$\,GeV. We can then integrate out the heavy SM particles -- the top quark, the Higgs boson and the weak gauge bosons $W^\pm$ and $Z^0$ -- at the scale $\mu_w$ and match the effective Lagrangian (\ref{LeffmuW}) onto a low-energy effective Lagrangian in which these degrees of freedom are no longer present as propagating fields. Just below the scale $\mu_w$, this Lagrangian takes the form
\begin{equation}\label{LlowE}
\begin{aligned}
   {\cal L}_{\rm eff}^{D\le 5}(\mu\lesssim \mu_w)
   &= \frac12 \left( \partial_\mu a\right)\!\left( \partial^\mu a\right) - \frac{m_{a,0}^2}{2}\,a^2
    + {\cal L}_{\rm ferm}'(\mu) \\
   &\quad\mbox{}+ c_{GG}\,\frac{\alpha_s}{4\pi}\,\frac{a}{f}\,G_{\mu\nu}^a\,\tilde G^{\mu\nu,a}
    + c_{\gamma\gamma}\,\frac{\alpha}{4\pi}\,\frac{a}{f}\,F_{\mu\nu}\,\tilde F^{\mu\nu} \,,
\end{aligned}
\end{equation}
where ${\cal L}_{\rm ferm}'$ is given by (\ref{Lferm}) but with the top-quark fields $t_L$ and $t_R$ removed. In general, the Wilson coefficients $c_{GG}$, $c_{\gamma\gamma}$, $\bm{k}_F$ and $\bm{k}_f$ in this effective Lagrangian differ from the corresponding coefficients in the effective Lagrangian above the weak scale by calculable matching contributions, which arise when the weak-scale particles are integrated out. We now discuss the calculation of the relevant matching conditions at one-loop and partial two-loop order. 

\subsection{Matching contributions to the ALP--boson couplings}

\begin{figure}
\begin{center}
\includegraphics[height=2.2cm]{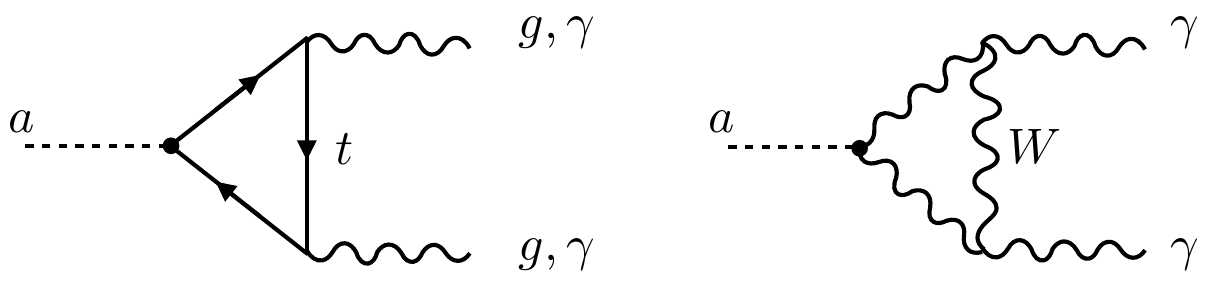}
\vspace{2mm}
\caption{\label{fig:matching} 
Examples of one-loop matching contributions to the ALP--boson couplings. These diagrams do not give rise to matching contributions when the form of the effective Lagrangian in (\ref{Leff}) is employed.} 
\end{center}
\end{figure}

One-loop matching corrections to the ALP--gluon and ALP--photon couplings $c_{GG}$ and $c_{\gamma\gamma}$ could in principle arise from loop graphs containing top quarks and heavy electroweak gauge bosons. Two representative diagrams are shown in Figure~\ref{fig:matching}. The corresponding effects were calculated in \cite{Bauer:2017ris}, and it was shown that for a light ALP these effects decouple like $m_a^2/m_t^2$ and $m_a^2/m_W^2$, respectively. For a light ALP far below the weak scale there are thus no matching contributions to the effective low-energy Lagrangian (\ref{LlowE}) from these loops, i.e.
\begin{equation}
   \Delta c_{GG}(\mu_w) = 0 \,, \qquad \Delta c_{\gamma\gamma}(\mu_w) = 0 \,.
\end{equation}
Matching corrections of order $m_a^2/m_t^2$ or $m_a^2/m_W^2$, which arise from the Taylor expansions of the functions $B_1(\tau)$ and $B_2(\tau)$ in (\ref{B2def}) in the region where $\tau\gg 1$, would contribute to the Wilson coefficients of dimension-7 operators in the low-energy effective theory below the weak scale, which we neglect for simplicity.

As a side remark, let us mention briefly that the situation would be different if we were to perform the calculations based on the alternative form of the effective Lagrangian shown in (\ref{Leffalt}). In this case there are non-vanishing matching contributions from top-quark loop diagrams, which lead to 
\begin{equation}\label{topmatch}
   \Delta\tilde c_{GG}(\mu_w) = - \frac{c_{tt}(\mu_w)}{2} \,, \qquad
   \Delta\tilde c_{\gamma\gamma}(\mu_w) = - \frac{4c_{tt}(\mu_w)}{3} \,.
\end{equation}
Recall that, according to (\ref{rela2}) and (\ref{eq:30}), the coefficients $\tilde c_{GG}$ and $\tilde c_{\gamma\gamma}$ above the weak scale are related to the corresponding unprimed coefficients by
\begin{equation}
\begin{aligned}
   \tilde c_{GG}(\mu>\mu_w) &= c_{GG} + \frac12\spac\sum_q\spac c_{qq}(\mu) 
    = \tilde c_{GG}(\mu) \,, \\
   \tilde c_{\gamma\gamma}(\mu>\mu_w) &= c_{\gamma\gamma} 
    + \sum_f\spac N_c^f\spac Q_f^2\,c_{ff}(\mu) = \tilde c_{\gamma\gamma}(\mu) \,,
\end{aligned}
\end{equation}
where the sum in the first (second) equation runs over all quark (fermion) species in the SM. When crossing the weak scale, one needs to add the matching contributions given above, and this has the effect of removing the contributions from the top quark in these relations. We thus obtain
\begin{equation}\label{cgagaeffdef}
\begin{aligned}
   \tilde c_{GG}(\mu\lesssim\mu_w) 
   &= c_{GG} + \frac12\spac\sum_q\spac c_{qq}(\mu) + \Delta\tilde c_{GG}(\mu_w)
    = c_{GG} + \frac12\spac\sum_{q\ne t}\spac c_{qq}(\mu) \,, \\
   \tilde c_{\gamma\gamma}(\mu\lesssim\mu_w) 
   &= c_{\gamma\gamma} + \sum_f\spac N_c^f\spac Q_f^2\,c_{ff}(\mu) + \Delta\tilde c_{\gamma\gamma}(\mu_w)
    = c_{\gamma\gamma} + \sum_{f\ne t}\spac N_c^f\spac Q_f^2\,c_{ff}(\mu) \,.
\end{aligned}
\end{equation}
The same procedure repeats itself as $\mu$ is evolved to lower energies and one crosses the threshold of other heavy fermions.

\subsection{Matching contributions to the ALP--fermion couplings}

\begin{figure}
\begin{center}
\includegraphics[height=2.1cm]{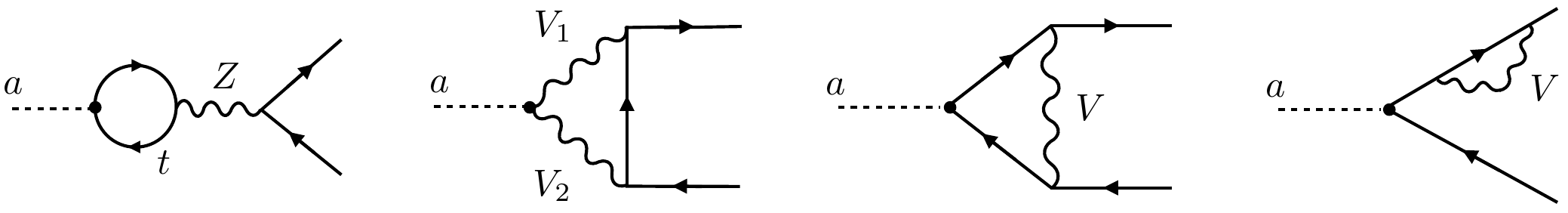}
\vspace{2mm}
\caption{\label{fig:graphs_ato2e} 
One-loop matching contributions to the ALP--fermion couplings. In the second diagram $(V_1 V_2)=(WW)$, $(ZZ)$, $(Z\gamma)$ or $(\gamma Z)$. In the last two diagrams $V=W,Z$, but in the sum of all contributions only the $W$-boson graphs with internal top quarks (plus the corresponding graphs with Goldstone bosons) give rise to non-zero contributions.} 
\end{center}
\end{figure}

One-loop matching corrections to the ALP--fermion couplings arise from graphs containing heavy electroweak gauge bosons. Some representative diagrams are shown in Figure~\ref{fig:graphs_ato2e}. Loop diagrams involving Higgs bosons give contributions proportional to the Yukawa couplings of the external fermions. Since the top quark is integrated out in the effective theory below the weak scale, these graphs are proportional to $y_f^2$ for some light SM fermion $f$ and hence can be neglected. The first diagram in Figure~\ref{fig:graphs_ato2e} arises from ALP mixing with the $Z$ boson via a top-quark loop. The second graph gives rise to matching contributions proportional to the ALP--boson couplings. The corresponding effects were calculated in \cite{Bauer:2017ris} for the case where the external fermions are leptons. Here we generalize these results to the case of quarks, where however contributions involving virtual top quarks require a special treatment. The remaining diagrams contain vertex and external-leg corrections from loops involving heavy $W$ and $Z$ bosons. We have calculated these diagrams in a general $R_\xi$ gauge, finding that the sum of all contributions yields a gauge-invariant answer. Moreover, the sum of all contributions involving $Z$ bosons and their Goldstone bosons vanishes. For the diagrams involving $W$ bosons a non-zero contribution remains, which arises from graphs containing internal top quarks. These diagrams contribute to the couplings $\bm{k}_D(\mu_w)$ in the left-handed down-quark sector only, and they are the only source of flavor off-diagonal effects. Combining all terms, we find the matching contributions (with $F=U,D,E,\nu$ and $f=u,d,e$) 
\begin{align}\label{cffmatching}
   \Delta\bm{k}_F(\mu_w) 
   &= \frac{3 y_t^2}{8\pi^2}\,c_{tt}\,\big( T_3^f - Q_f\spac s_w^2 \big)\,
    \ln\frac{\mu_w^2}{m_t^2}\,\mathbbm{1} \notag\\
   &\quad\mbox{}+ \frac{3\alpha^2}{8\pi^2}\,\bigg[    
    \frac{c_{WW}}{2s_w^4} \left( \ln\frac{\mu_w^2}{m_W^2} + \frac12 + \delta_1 \right) 
    + \frac{2c_{\gamma Z}}{s_w^2\spac c_w^2}\,Q_f\,\big( T_3^f - Q_f\spac s_w^2 \big)
    \left( \ln\frac{\mu_w^2}{m_Z^2} + \frac32 + \delta_1 \right) \notag\\
   &\hspace{1.5cm}\mbox{}+ \frac{c_{ZZ}}{s_w^4\spac c_w^4}\,\big( T_3^f - Q_f\spac s_w^2 \big)^2
    \left( \ln\frac{\mu_w^2}{m_Z^2} + \frac12 + \delta_1 \right) \!\bigg]\,\mathbbm{1} 
    + \delta_{FD}\,\hat\Delta\bm{k}_D(\mu_w) \,, \notag\\
   \Delta\bm{k}_f(\mu_w) 
   &= \frac{3 y_t^2}{8\pi^2}\,c_{tt}\,\big(\! - Q_f\spac s_w^2 \big)\,
    \ln\frac{\mu_w^2}{m_t^2}\,\mathbbm{1} \notag\\
   &\quad\mbox{}+ \frac{3\alpha^2}{8\pi^2}\,Q_f^2 \left[
    \frac{2c_{\gamma Z}}{c_w^2} \left( \ln\frac{\mu_w^2}{m_Z^2} + \frac32 + \delta_1 \right) 
    - \frac{c_{ZZ}}{c_w^4} \left( \ln\frac{\mu_w^2}{m_Z^2} + \frac12 + \delta_1 \right) \right] 
    \mathbbm{1} \,.
\end{align}
These contributions must be added to the RG-evolved coefficients at $\mu=\mu_w$, so that one obtains $\bm{k}_{F,f}(\mu_w)+\Delta\bm{k}_{F,f}(\mu_w)$ for the ALP--fermion couplings just below the weak scale. All scale-dependent parameters on the right-hand side of the above relations are evaluated at the scale $\mu_w$. RG invariance requires that the ALP--boson couplings entering in these relations must appear in the form of the couplings $\tilde c_{V_1 V_2}$, at least in the coefficients of the $\ln(\mu_w^2/m_{W,Z}^2)$ terms. Hence, via the substitution $c_{V_1 V_2}\to\tilde c_{V_1 V_2}$ we can account for an important subclass of two-loop matching contributions. The scheme-dependent constant $\delta_1$ arises from the treatment of the Levi--Civita symbol in $d$ dimensions. We obtain $\delta_1=-\frac{11}{3}$ in a scheme where $\epsilon^{\mu\nu\alpha\beta}$ is treated as a $d$-dimensional object, and $\delta_1=0$ if it is instead treated as a 4-dimensional quantity. 

The non-trivial flavor structure is captured by the quantity
\begin{align}\label{eq:KDeff}
   \big[ \hat\Delta k_D(\mu_w) \big]_{ij}
   &= \frac{y_t^2}{16\pi^2}\,\bigg\{
    V_{mi}^* V_{nj} \left[ k_U(\mu_w) \right]_{mn} \left( \delta_{m3} + \delta_{n3} \right)
    \left[ - \frac14\ln\frac{\mu_w^2}{m_t^2} - \frac38
    + \frac34\,\frac{1-x_t+\ln x_t}{\left(1-x_t\right)^2} \right] \notag\\
   &\hspace{1.7cm}\mbox{}+ V_{3i}^* V_{3j} \left[ k_U(\mu_w) \right]_{33} 
    + V_{3i}^* V_{3j} \left[ k_u(\mu_w) \right]_{33}\! \left[ 
    \frac12\ln\frac{\mu_w^2}{m_t^2} - \frac14 
    - \frac32\,\frac{1-x_t+\ln x_t}{\left(1-x_t\right)^2} \right] \notag\\
   &\hspace{1.7cm}\mbox{}- \frac{3\alpha}{2\pi s_w^2}\,c_{WW}\,V_{3i}^* V_{3j}\,
    \frac{1-x_t+x_t\ln x_t}{\left(1-x_t\right)^2} \bigg\} \,, 
\end{align}
where $x_t=m_t^2/m_W^2$. These matching contributions are sources of flavor-changing ALP interactions even if the underlying UV theory does not contain new sources of flavor or CP violation beyond those present in the SM. We have neglected the Yukawa couplings of the light quarks and leptons. In this approximation there are no flavor off-diagonal matching contributions in the up-quark and lepton sectors. 

\subsection{ALP--fermion couplings below the electroweak scale}

\subsubsection*{Flavor-diagonal couplings}

The flavor-diagonal ALP--fermion interactions in (\ref{Lferm}) can be expressed in terms of vector and axial-vector currents. The vector currents are conserved below the weak scale and thus do not contribute to physical matrix elements. It follows that we can rewrite this Lagrangian in the equivalent form (for $\mu\lesssim\mu_w$)
\begin{equation}\label{eqcff}
   {\cal L}_{\rm ferm}^{\rm diag}(\mu) 
   = \sum_{f\ne t}\,\frac{c_{ff}(\mu)}{2}\,\frac{\partial^\mu a}{f}\,
    \bar f\spac\gamma_\mu\gamma_5\spac f \,,
\end{equation}
where the sum runs over all charged fermion species in the low-energy theory (the quarks $u,d,s,c,b$ and the leptons $e,\mu,\tau$). The couplings $c_{ff}$ have been defined in (\ref{cffdef}) in terms of the diagonal elements of the matrices $\bm{k}_f$ and $\bm{k}_F$. Note that the ALP--neutrino interactions can be dropped in the low-energy Lagrangian (but not in the theory above the weak scale, where they contribute at one-loop order to the ALP couplings to $W$ and $Z$ bosons). Using integration by parts, the derivative on the neutrino axial-vector current vanishes because the neutrinos are massless in the SM. 

At the matching scale $\mu_w$, the coefficients $c_{ff}(\mu_w)$ are given by the sum of the contributions from RG evolution, shown in (\ref{eq:51}) and (\ref{eq:52}), and weak-scale matching, see (\ref{cffmatching}) and (\ref{eq:KDeff}). In this sum the dependence on the matching scale $\mu_w$ partially cancels out; however, some scale dependence remains and cancels when the evolution below the weak scale is taken into account (see Section~\ref{subsec:RGElow} below). In order to get a feeling for the magnitude of the radiative corrections we choose the new-physics scale $\Lambda=4\pi f$ with $f=1$\,TeV and evaluate the coefficients $c_{ff}(\mu)$ in the vicinity of $\mu_w=m_t$. We find numerically 
\begin{align}\label{cffrun}
   c_{uu,cc}(\mu_w) 
   &\simeq c_{uu,cc}(\Lambda) - 0.116\,c_{tt}(\Lambda) 
    - \Big[ 6.35\,\tilde c_{GG}(\Lambda) + 0.19\,\tilde c_{WW}(\Lambda) 
    + 0.02\,\tilde c_{BB}(\Lambda) \Big]\cdot 10^{-3} \notag\\
   &\quad\mbox{}- \left[ \tilde c_{GG}(\mu_w)\,\frac{\alpha_s^2(\mu_w)}{\pi^2}
    + \tilde c_{\gamma\gamma}(\mu_w)\,\frac{\alpha^2(\mu_w)}{3\pi^2} \right] 
    \ln\frac{m_t^2}{\mu_w^2} \,, \notag\\
   c_{dd,ss}(\mu_w) 
   &\simeq c_{dd,ss}(\Lambda) + 0.116\,c_{tt}(\Lambda) 
    - \Big[ 7.08\,\tilde c_{GG}(\Lambda) + 0.22\,\tilde c_{WW}(\Lambda) 
    + 0.005\,\tilde c_{BB}(\Lambda) \Big]\cdot 10^{-3} \notag\\
   &\quad\mbox{}- \left[ \tilde c_{GG}(\mu_w)\,\frac{\alpha_s^2(\mu_w)}{\pi^2}
    + \tilde c_{\gamma\gamma}(\mu_w)\,\frac{\alpha^2(\mu_w)}{12\pi^2} \right] 
    \ln\frac{m_t^2}{\mu_w^2} \,, \notag\\
   c_{bb}(\mu_w) 
   &\simeq c_{bb}(\Lambda) + 0.097\,c_{tt}(\Lambda) 
    - \Big[ 7.02\,\tilde c_{GG}(\Lambda) + 0.19\,\tilde c_{WW}(\Lambda)
    + 0.005\,\tilde c_{BB}(\Lambda) \Big] \cdot 10^{-3} \notag\\
   &\quad\mbox{}- \left[ \tilde c_{GG}(\mu_w)\,\frac{\alpha_s^2(\mu_w)}{\pi^2}
    + \tilde c_{\gamma\gamma}(\mu_w)\,\frac{\alpha^2(\mu_w)}{12\pi^2} \right] 
    \ln\frac{m_t^2}{\mu_w^2} \,, \notag\\   
   c_{e_i e_i}(\mu_w) 
   &\simeq c_{e_i e_i}(\Lambda) + 0.116\,c_{tt}(\Lambda) 
    - \Big[ 0.37\,\tilde c_{GG}(\Lambda) + 0.22\,\tilde c_{WW}(\Lambda) 
    + 0.05\,\tilde c_{BB}(\Lambda) \Big]\cdot 10^{-3} \notag\\
   &\quad\mbox{}- \tilde c_{\gamma\gamma}(\mu_w)\,
    \frac{3\alpha^2(\mu_w)}{4\pi^2}\,\ln\frac{m_t^2}{\mu_w^2} \,.
\end{align}
We use the two-loop expression for the running coupling $\alpha_s(\mu)$ and the one-loop approximations for the couplings $\alpha_1(\mu)$ and $\alpha_2(\mu)$, and we evaluate the function $U(\mu_w,\Lambda)$ using the explicit form (\ref{Usolu}). For the couplings $c_{d_i d_i}$ in the down-quark sector we work under the assumption of minimal flavor violation and have approximated $|V_{tb}|^2\approx 1$ and $|V_{td}|^2\approx|V_{ts}|^2\approx 0$. From (\ref{eq:20}), the matching conditions $\tilde c_{VV}(\Lambda)$ can be written in the form
\begin{equation}
\begin{aligned}
   \tilde c_{GG}(\Lambda) 
   &= c_{GG} + \frac12\spac\sum_q\spac c_{qq}(\Lambda) \,, \\[-2mm]
   \tilde c_{WW}(\Lambda) 
   &= c_{WW} - \frac12\,\text{Tr}\spac\Big[ 3\bm{k}_U(\Lambda) + \bm{k}_E(\Lambda) \Big] \,, \\[1mm]
   \tilde c_{BB}(\Lambda) 
   &= c_{BB} + \sum_f\spac N_c^f\spac Q_f^2\,c_{ff}(\Lambda) 
    + \frac12\,\text{Tr}\spac\Big[ 3\bm{k}_U(\Lambda) + \bm{k}_E(\Lambda) \Big] \,,
\end{aligned}
\end{equation}
where the sums run over all quark and fermion flavors. We observe that electroweak radiative corrections are generally very small, while the contributions proportional to $c_{tt}$ from the Yukawa interactions as well as QCD effects can be sizable. For example, in scenarios where the ALP--boson couplings at the UV scale are an order of magnitude larger than the ALP-fermion couplings, the corrections induced by $\tilde c_{GG}$ can give contributions to $c_{qq}(\mu_w)$ of about 7\%, whereas the contributions of $\tilde c_{WW}$ and $\tilde c_{BB}$ are negligible. The logarithms of the ratio $(m_t^2/\mu_w^2)$ in the above expressions show the remaining dependence on the weak matching scale. This dependence cancels out when evolution effects below the weak scale are included. 

The numerical results shown in (\ref{cffrun}) are relevant for an ALP which is part of a new-physics sector at a scale $\Lambda\sim 10$\,TeV. For the QCD axion, one typically considers much higher scales in the vicinity of $f\sim 10^{12\pm 3}$\,GeV. This gives rise to significantly enhanced evolution effects. For example, choosing $\Lambda=4\pi f$ with $f=10^{12}$\,GeV and setting $\mu_w=m_t$ we find 
\begin{equation}
\begin{aligned}
   c_{uu,cc}(m_t) 
   &\simeq c_{uu,cc}(\Lambda) - 0.350\,c_{tt}(\Lambda) 
    - \Big[ 12.6\,\tilde c_{GG}(\Lambda) + 0.84\,\tilde c_{WW}(\Lambda) 
    + 0.10\,\tilde c_{BB}(\Lambda) \Big]\cdot 10^{-3} \,, \\
   c_{dd,ss}(m_t) 
   &\simeq c_{dd,ss}(\Lambda) + 0.353\,c_{tt}(\Lambda) 
    - \Big[ 16.8\,\tilde c_{GG}(\Lambda) + 1.30\,\tilde c_{WW}(\Lambda) 
    + 0.07\,\tilde c_{BB}(\Lambda) \Big]\cdot 10^{-3} \,, \\
   c_{bb}(m_t) 
   &\simeq c_{bb}(\Lambda) + 0.294\,c_{tt}(\Lambda) 
    - \Big[ 16.5\,\tilde c_{GG}(\Lambda) + 1.23\,\tilde c_{WW}(\Lambda)
    + 0.06\,\tilde c_{BB}(\Lambda) \Big] \cdot 10^{-3} \,, \\   
   c_{e_i e_i}(m_t) 
   &\simeq c_{e_i e_i}(\Lambda) + 0.352\,c_{tt}(\Lambda) 
    - \Big[ 2.09\,\tilde c_{GG}(\Lambda) + 1.30\,\tilde c_{WW}(\Lambda) 
    + 0.38\,\tilde c_{BB}(\Lambda) \Big]\cdot 10^{-3} \,.
\end{aligned}
\end{equation}
The contributions proportional to $c_{tt}$ now give ${\cal O}(1)$ corrections to all ALP--fermion couplings.

It is very useful to derive a simple, approximate expression for the ALP--fermion couplings at the scale $\mu_w$, in which one neglects the small two-loop electroweak evolution effects as well as the two-loop contributions proportional to the ALP--fermion couplings themselves. This yields (for $q\ne t$ and $\mu\lesssim\mu_w$)
\begin{align}\label{boundlessbeauty}
   c_{qq}(\mu) &\approx c_{qq}(\Lambda) 
    - 6\spac T_3^q \left( 1 - \frac{\delta_{qb}}{6} \right) \frac{\alpha_t(m_t)}{\alpha_s(m_t)}\,
    \bigg[ 1 - \left( \frac{\alpha_s(\Lambda)}{\alpha_s(m_t)} \right)^{\!\frac17} \bigg]\,c_{tt}(\Lambda) 
    - \frac{4 c_{GG}}{\beta_0^{(3)}}\,\frac{\alpha_s(\mu)-\alpha_s(\Lambda)}{\pi} \,, \notag\\
   c_{\ell\ell}(\mu) &\approx c_{\ell\ell}(\Lambda) + 3\,\frac{\alpha_t(m_t)}{\alpha_s(m_t)}\,
    \bigg[ 1 - \left( \frac{\alpha_s(\Lambda)}{\alpha_s(m_t)} \right)^{\!\frac17} \bigg]\,
    c_{tt}(\Lambda) \,,
\end{align}
which as we will see in the next section continues to hold below the weak scale. Note that only the last term in the first line is scale-dependent in this approximation, and one needs to adjust the value of $\beta_0^{(3)}$ whenever one crosses a quark threshold. In the first relation $T_3^u=\frac12$ and $T_3^d=-\frac12$ denotes the weak isospin. In the above expressions large logarithms of the scale ratio $\Lambda/\mu_w$ are resummed to all orders in perturbation theory. The most striking effect is the universal admixture (weighted only by weak isospin) of a contribution proportional to $c_{tt}(\Lambda)$ to all ALP--fermion couplings, even those involving the charged leptons. When one re-expands the above expressions to first order in couplings, one obtains
\begin{equation}\label{approx}
   c_{ff}(\mu) 
   \approx c_{ff}(\Lambda) - \left( 1 - \frac{\delta_{fb}}{6} \right)
    \frac{3\spac y_t^2(m_t)}{8\pi^2}\,T_3^f\,c_{tt}(\Lambda)\,\ln\frac{\Lambda^2}{m_t^2} \,.
\end{equation}
This effect was noted previously in \cite{MartinCamalich:2020dfe}, where the opposite sign was obtained and in the argument of the logarithm the scale $\mu$ was used rather than $m_t$. Note, however, that this effect is due to the first diagram in Figure~\ref{fig:graphs_ato2e}, which no longer contributes below the scale of the top quark. Also, the resummation effects included here can be numerically very important. With $f=10^{12}$\,GeV, for instance, formula (\ref{approx}) would predict $\pm 0.84\,c_{tt}(\Lambda)$, overshooting the effect by more than a factor of 2.

\begin{figure}
\centering
\includegraphics[height=6.5cm]{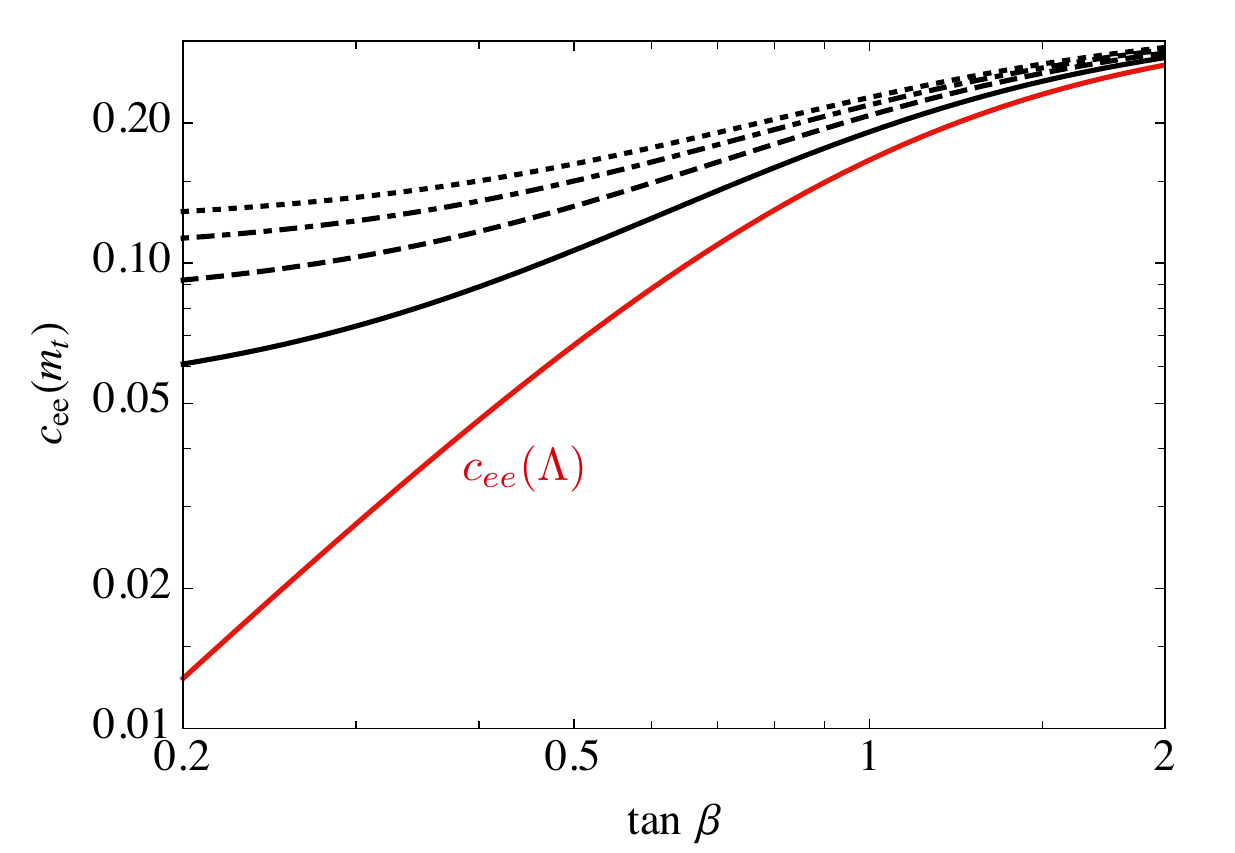}
\caption{\label{fig:cee} 
Axion--electron coupling $c_{ee}(m_t)$ in the DFSZ model for different values of $\tan\beta=v_u/v_d$ and axion masses: $m_a=1$\,keV (solid), 1\,eV (dashed), 1\,meV (dashed-dotted) and $1\,\mu$eV (dotted). The red curve depicts the coupling $c_{ee}(\Lambda)$ at the high scale $\Lambda=4\pi f$.}
\end{figure}

The evolution effects in (\ref{boundlessbeauty}) are of potentially large importance not only for ALPs, but also for the classical QCD axion. In order to illustrate this fact we consider the DFSZ model \cite{Dine:1981rt,Zhitnitsky:1980tq}, in which the ALP couplings at the UV scale $\Lambda=4\pi f$ satisfy \cite{Srednicki:1985xd,Zyla:2020zbs}
\begin{equation}
   c_{u_i u_i}(\Lambda) = \frac13 \cos^2\beta \,, \qquad
   c_{d_i d_i}(\Lambda) = c_{e_i e_i}(\Lambda) = \frac13 \sin^2\beta \,, \qquad
   c_{GG} = - \frac12 \,,
\end{equation}
where $\tan\beta=v_u/v_d$ is the ratio of the vacuum expectation values of the two Higgs doublets, with a phenomenologically motivated range spanning $0.28<\tan\beta<140$ \cite{Chen:2013kt}. The axion mass is given by relation (\ref{massrela}) with $m_{a,0}^2=0$, i.e.\ it is uniquely determined by the decay constant $f$. Assuming that the masses of the additional Higgs bosons are larger than $\Lambda$, we can evolve these coupling parameters down to the weak scale. Figure~\ref{fig:cee} shows the axion--electron coupling at the high scale (red line) and the RG-evolved couplings $c_{ee}(m_t)$ at the electroweak scale for different axion masses. The smaller the axion mass, the larger are the evolution effects because the corresponding values of $\Lambda$ increase proportional to $1/m_a$, ranging from $\Lambda\simeq 73$\,TeV for $m_a=1$\,keV to $\Lambda\simeq 7.3\cdot 10^{10}$\,TeV for $m_a=1\,\mu$eV. The figure shows that in the DFSZ model the axion--electron coupling can be enhanced through evolution effects by up to an order of magnitude for small values of $\tan\beta$.

\subsubsection*{Flavor-changing couplings}

The flavor-changing ALP--fermion couplings in (\ref{Lferm}) can be integrated by parts without introducing additional contributions to the Wilson  coefficients $c_{VV}$. This gives (for $\mu\lesssim\mu_w$)
\begin{equation}\label{masssuppression}
\begin{aligned}
   {\cal L}_{\rm ferm}^{\rm FCNC}(\mu) 
   &= - \frac{ia}{2f}\,\sum_f\,\Big[ 
    (m_{f_i}-m_{f_j}) \left( k_f+k_F \right)_{ij} \bar f_i\spac f_j 
    + (m_{f_i}+m_{f_j}) \left( k_f-k_F \right)_{ij} \bar f_i\spac\gamma_5 f_j \Big] \,,
\end{aligned}
\end{equation}
where throughout this discussion $i\ne j$. The fermion masses and coupling parameters must be evaluated at the scale $\mu$. This form of the Lagrangian makes explicit that flavor-changing amplitudes are suppressed by the masses of the fermions involved. (The same is true for the flavor-conserving interactions in (\ref{eqcff}), but in this case integrating by parts generates additional contributions to the ALP--gluon and ALP--photon couplings.) At the weak scale $\mu_w$, the generation off-diagonal coefficients $\left[k_f(\mu_w)\right]_{ij}$ and $\left[k_F(\mu_w)\right]_{ij}$ are again given by the sum of the contributions from RG evolution and weak-scale matching. Recall that generation off-diagonal matching contributions are captured by the quantity $\hat\Delta\bm{k}_D(\mu_w)$ in (\ref{eq:KDeff}). For all coefficients other than $\bm{k}_D$, one finds from (\ref{cijsolu}) and (\ref{cQ33cQ11}) that flavor-changing interactions at the weak scale are inherited from the UV scale $\Lambda$. We find 
\begin{equation}\label{eq:47}
\begin{aligned}
   \left[ k_u(\mu_w) \right]_{ij} 
   &= \left[ k_u(\Lambda) \right]_{ij} ; \quad i,j\ne 3 \,, \\
   \left[ k_U(\mu_w) \right]_{ij} 
   &= \left[ k_U(\Lambda) \right]_{ij} ; \quad i,j\ne 3 \,, \\
   \left[ k_d(\mu_w) \right]_{ij} 
   &= \left[ k_d(\Lambda) \right]_{ij} , \\
   \left[ k_e(\mu_w) \right]_{ij}
   &= \left[ k_e(\Lambda) \right]_{ij} , \\
   \left[ k_E(\mu_w) \right]_{ij} 
   &= \left[ k_E(\Lambda) \right]_{ij} .
\end{aligned}
\end{equation}
Note that for $\bm{k}_u$ and $\bm{k}_U$ we only need the entries where $i,j\ne 3$, since the top quark has been integrated out in the effective theory below the weak scale. If the UV theory respects minimal flavor violation, then all these couplings vanish. For the off-diagonal elements of the coefficient $\bm{k}_D$ we find the more interesting result
\begin{equation}\label{eq:48}
\begin{aligned}
   \big[ k_D(\mu_w) \big]_{ij}
   &= \big[ \bm{V}^\dagger \bm{k}_U(\Lambda) \bm{V} \big]_{ij} 
    - V_{mi}^* V_{nj} \left( \delta_{m3} + \delta_{n3} - 2\spac\delta_{m3}\spac\delta_{n3} \right) 
    \left( 1 - e^{-U(\mu_w,\Lambda)} \right) \left[ k_U(\Lambda) \right]_{mn} \\
   &\quad\mbox{}- \frac16\,V_{3i}^* V_{3j}\,I_t(\mu_w,\Lambda) 
    + \big[ \hat\Delta k_D(\mu_w) \big]_{ij} \,,
\end{aligned}
\end{equation}
where the integral $I_t(\mu_w,\Lambda)$ has been defined in (\ref{Itdef}). If the original ALP Lagrangian (\ref{Leff}) at the new-physics scale respects the principle of minimal flavor violation, the matrix $\bm{k}_U$ is diagonal, as shown in (\ref{MFV}). In this case the above expression simplifies significantly, and we find 
\begin{equation}\label{sonice}
\begin{aligned}
   \big[ k_D(\mu_w) \big]_{ij}
   &=  V_{ti}^* V_{tj}\,\bigg\{ 
    \left[ (k_U)(\Lambda) \right]_{33} - \left[ (k_U)(\Lambda) \right]_{11}
    - \frac16\,I_t(\mu_w,\Lambda) \\
   &\hspace{1.9cm}\mbox{}+ \frac{y_t^2(\mu_w)}{16\pi^2}\,\bigg[ c_{tt}(\mu_w) 
    \left[ \frac12\ln\frac{\mu_w^2}{m_t^2} - \frac14 
    - \frac32\,\frac{1-x_t+\ln x_t}{\left(1-x_t\right)^2} \right] \\
   &\hspace{3.9cm}\mbox{}- \frac{3\alpha}{2\pi\spac s_w^2}\,c_{WW}\,
    \frac{1-x_t+x_t\ln x_t}{\left(1-x_t\right)^2} \bigg] \bigg\} \,,
\end{aligned}
\end{equation}
where (again setting the new-physics scale to $\Lambda=4\pi f$ with $f=1$\,TeV)
\begin{equation}
   c_{tt}(\mu_w) \simeq 0.826\,c_{tt}(\Lambda) 
    - \big[ 6.17\,\tilde c_{GG}(\Lambda) + 0.23\,\tilde c_{WW}(\Lambda)
    + 0.02\,\tilde c_{BB}(\Lambda) \big]\cdot 10^{-3} \,.
\end{equation}
Note that under the hypothesis of minimal flavor violation the matrix $\bm{k}_U$ is diagonal but not necessarily proportional to the unit matrix in generation space, see (\ref{MFV}). The first term on the right-hand side of (\ref{sonice}) thus accounts for the possibility that $\left[(k_U)(\Lambda)\right]_{33}\ne\left[(k_U)(\Lambda)\right]_{11}$. If this is the case, then the off-diagonal matrix elements 
\begin{equation}
   \big[ k_D(\Lambda) \big]_{ij}
   =  V_{ti}^* V_{tj}\,\Big\{ \left[ (k_U)(\Lambda) \right]_{33}
    - \left[ (k_U)(\Lambda) \right]_{11} \!\Big\}
\end{equation}
at the new-physics scale can be non-zero, providing a UV source of flavor violation. Evolving the coefficients to the weak scale $\mu_w=m_t$, we obtain numerically 
\begin{equation}\label{eq:FVveryshort}
   \left[ k_D(m_t) \right]_{ij}
   \simeq \left[ k_D(\Lambda) \right]_{ij} 
    + 0.019\,V_{ti}^* V_{tj}\,\Big[ c_{tt}(\Lambda) 
    - 0.0032\,\tilde c_{GG}(\Lambda) - 0.0057\,\tilde c_{WW}(\Lambda) \Big] \,.
\end{equation}
The matching contributions proportional to $\tilde c_{GG}$ and $\tilde c_{WW}$ are very small. 

Relation (\ref{sonice}) shows explicitly how flavor-changing effects are generated through RG evolution from the new-physics scale $\Lambda$ to the weak scale (first line) and matching contributions at the weak scale (second and third lines). These loop-induced effects should be considered as the minimal effects of flavor violation present in any ALP model, even if the matrix $\bm{k}_D$ is diagonal at the new-physics scale $\Lambda$ (which would be a stronger assumption than minimal flavor violation). The terms proportional to $c_{WW}$ in (\ref{sonice}) agree with a corresponding expression derived in \cite{Izaguirre:2016dfi}. Our results for the evolution effects and the contribution proportional to $c_{tt}(\mu_w)$ are new. The logarithm of $(\mu_w^2/m_t^2)$ in the coefficient of $c_{tt}$ (but not the $x_t$-dependent remainder) was derived in \cite{Gavela:2019wzg}. The more general expressions shown above, and in particular the results (\ref{eq:KDeff}) and (\ref{eq:48}), which do not assume minimal flavor violation, are derived here for the first time.

In the sum of the contributions from scale evolution and weak-scale matching, the dependence on the matching scale $\mu_w$ drops out. This is obviously true for the coefficients in (\ref{eq:47}), but it also holds for the sum of all terms on the right-hand side of (\ref{eq:48}). In fact, we will see in Section~\ref{subsec:RGElow} that the flavor off-diagonal Wilson coefficients do not run below the weak scale (in the approximation where the Yukawa couplings of the light quarks are put to zero). Hence, the expressions shown in (\ref{eq:47}) and (\ref{eq:48}) hold for all values $\mu<\mu_w$.

\section{Renormalization-group evolution below the weak scale}
\label{subsec:RGElow}

Now that we have obtained the values of the Wilson coefficients at the weak scale, we should evolve these coefficients down to lower scales, so that they can be used in calculations of low-energy observables. Compared with (\ref{RGEs}) the evolution equations simplify significantly, because the Yukawa interactions mediated by Higgs exchange are absent in the low-energy theory, as are diagrams including the heavy weak gauge bosons. The only remaining contributions to the evolution equations result from the second diagram in Figure~\ref{fig:2loop} and the last diagram in Figure~\ref{fig:mixing}, where the gauge bosons can be gluons or photons. We obtain 
\begin{equation}\label{eq:53}
\begin{aligned}
   \frac{d}{d\ln\mu}\,\bm{k}_q(\mu) = - \frac{d}{d\ln\mu}\,\bm{k}_Q(\mu) 
   &= \left( \frac{\alpha_s^2}{\pi^2}\,\tilde c_{GG} 
    + \frac{3\alpha^2}{4\pi^2}\,Q_q^2\,\tilde c_{\gamma\gamma} \right) \mathbbm{1} \,, \\ 
   \frac{d}{d\ln\mu}\,\bm{k}_e(\mu) = - \frac{d}{d\ln\mu}\,\bm{k}_E(\mu) 
   &= \frac{3\alpha^2}{4\pi^2}\,\tilde c_{\gamma\gamma}\,\mathbbm{1} \,, 
\end{aligned}
\end{equation}
where $Q=U,D$ and $q=u,d$. Below the weak scale the scale dependence of the effective coefficients $\tilde c_{GG}$ and $\tilde c_{\gamma\gamma}$ is very weak, since it only arises at two-loop order. At next-to-leading logarithmic order, it is consistent to neglect this effect. Note also that the evolution effects below the weak scale are diagonal in generation space, and hence the flavor-changing couplings are scale-independent in the low-energy theory, as stated above. For the flavor-diagonal couplings only the parameters $c_{ff}$ defined in (\ref{cffdef}) are physical. At next-to-leading logarithmic order, their scale evolution is given by 
\begin{equation}\label{lowEevolution}
\begin{aligned}
   c_{qq}(\mu) &= c_{qq}(\mu_w) 
    - \frac{4\tilde c_{GG}(\mu_w)}{\beta_0^{\rm QCD}}\,\frac{\alpha_s(\mu)-\alpha_s(\mu_w)}{\pi} 
    - Q_q^2\,\frac{3 \tilde c_{\gamma\gamma}(\mu_w)}{\beta_0^{\rm QED}}\,
    \frac{\alpha(\mu)-\alpha(\mu_w)}{\pi} \,, \\
   c_{\ell\ell}(\mu) &= c_{\ell\ell}(\mu_w) 
    - \frac{3 \tilde c_{\gamma\gamma}(\mu_w)}{\beta_0^{\rm QED}}\,
    \frac{\alpha(\mu)-\alpha(\mu_w)}{\pi} \,.
\end{aligned}
\end{equation}
In the low-energy theory below the weak scale the relevant $\beta$-function coefficients are $\beta_0^{\rm QCD}=11-\frac23\,n_q$ for QCD and $\beta_0^{\rm QED}=-\frac43\sum_f N_c^f Q_f^2$ for QED. Here $n_q$ denotes the number of light quark flavors with mass below the scale $\mu$, and the sum over $f$ includes all light fermions with mass below $\mu$. Note that the dependence on the matching scale $\mu_w$ cancels when the expressions given in (\ref{cffrun}) are used in the above relations.

According to (\ref{cgagaeffdef}), the effective Wilson coefficients $\tilde c_{GG}(\mu_w)$ and $\tilde c_{\gamma\gamma}(\mu_w)$ contain the $c_{ff}$ parameters of all light fermions in the effective theory below the scale $\mu_w$. Generalizing these results to lower scales, we define
\begin{equation}\label{eq:60}
\begin{aligned}
   \tilde c_{GG}(\mu) &= c_{GG} + \frac12\,\sum_q\,c_{qq}(\mu)\,\theta(\mu-m_q) \,, \\
   \tilde c_{\gamma\gamma}(\mu) &= c_{\gamma\gamma} 
    + \sum_f\,N_c^f\,Q_f^2\,c_{ff}(\mu)\,\theta(\mu-m_f) \,.
\end{aligned}
\end{equation}
Like the $\beta$-function coefficients $\beta_0^{\rm QCD}$ and $\beta_0^{\rm QED}$, the effective couplings change by discrete amounts whenever one crosses a flavor threshold, and an appropriate matching must be performed in the usual way. In other words, one first evolves the coefficients from the weak scale to the scale $\mu_b\simeq m_b$, then eliminates the bottom quark from the list of light fermions, then evolves from the $b$-quark scale to the scale $\mu_\tau\simeq m_\tau$, then eliminates the $\tau$ lepton from the list of light fermions, and so on. In each step the coefficients of the $\beta$-functions as well as the values of $\tilde c_{GG}$ and $\tilde c_{\gamma\gamma}$ need to be adjusted. Concretely, at values of $\mu$ just below the scale $\mu_b\sim m_b$, we obtain
\begin{equation}
\begin{aligned}
   c_{qq}(\mu\lesssim\mu_b) &= c_{qq}(\mu_w) 
    - \frac{4\tilde c_{GG}(\mu_w)}{\beta_0^{\rm QCD}}\,\frac{\alpha_s(\mu_b)-\alpha_s(\mu_w)}{\pi} 
    - Q_q^2\,\frac{3 \tilde c_{\gamma\gamma}(\mu_w)}{\beta_0^{\rm QED}}\,
    \frac{\alpha(\mu_b)-\alpha(\mu_w)}{\pi} \\
   &\quad\mbox{}- \frac{4\tilde c_{GG}(\mu_b)}{\beta_0^{\rm QCD}}\,
    \frac{\alpha_s(\mu)-\alpha_s(\mu_b)}{\pi} 
    - Q_q^2\,\frac{3 \tilde c_{\gamma\gamma}(\mu_b)}{\beta_0^{\rm QED}}\,
    \frac{\alpha(\mu)-\alpha(\mu_b)}{\pi} \,,
\end{aligned}
\end{equation}
and similarly for $c_{\ell\ell}(\mu)$. In the two last terms of the first line the ALP--boson couplings and the $\beta$-function coefficients are evaluated with $n_q=5$ active quark flavors, whereas in the second line they are evaluated with $n_q=4$ flavors. The numerical impact of these low-scale evolution effects is very small. For example, with $\mu_w=m_t$ and $\mu_0=2$\,GeV we find
\begin{equation}
\begin{aligned}
   c_{qq}(\mu_0) 
   &= c_{qq}(m_t) + \Big[ 3.0\,\tilde c_{GG}(\Lambda) - 1.4\,c_{tt}(\Lambda)
    - 0.6\,c_{bb}(\Lambda) \Big] \cdot 10^{-2} \\
   &\quad\mbox{}+ Q_q^2\,\Big[ 3.9\,\tilde c_{\gamma\gamma}(\Lambda) 
    - 4.7\spac c_{tt}(\Lambda) - 0.2\,c_{bb}(\Lambda) \Big] \cdot 10^{-5} \,, \\
   c_{\ell\ell}(\mu_0) 
   &= c_{\ell\ell}(m_t) + \Big[ 3.9 \,\tilde c_{\gamma\gamma}(\Lambda)
    - 4.7\spac c_{tt}(\Lambda) - 0.2\,c_{bb}(\Lambda) \Big] \cdot 10^{-5} \,.
\end{aligned}    
\end{equation}

It is instructive to compare the above results with analogous expressions derived for the quark coefficients $c_{qq}$ in \cite{diCortona:2015ldu}. In this paper only QCD evolution effects were included. The results obtained there are in agreement with our findings when we ignore the terms proportional to the electromagnetic coupling $\alpha$ in the first line of (\ref{lowEevolution}). However, in \cite{diCortona:2015ldu} the same equation was used to account for evolution effects above the electroweak scale. This ignores the by far dominant contributions from the top-quark Yukawa interactions in (\ref{boundlessbeauty}), which as we have discussed have an important impact on all ALP--fermion couplings.

The scale-dependent ALP--boson couplings $\tilde c_{VV}$ defined in (\ref{eq:60}) are not only relevant in the context of the evolution equations for the ALP--fermion couplings, but they are also closely related to some observables of phenomenological interest. In (\ref{aggrate}) and (\ref{rates}) we have given explicit expressions for the $a\to gg$ and $a\to\gamma\gamma$ decay rates, the latter of which plays a pivotal role in the phenomenology of a light ALP. The fermion loop function entering these expressions satisfies $B_1(\tau)\approx 1$ for $\tau\ll 1$ (corresponding to ``light'' fermions with $m_f\ll m_a$) and $B_1(\tau)\approx 0$ for $\tau\gg 1$ (corresponding to ``heavy'' fermions with $m_f\gg m_a$). Moreover, the loop function $B_2(4m_W^2/m_a^2)\approx 0$ for a light ALP with mass $m_a\ll m_W$. Let us now apply an $\overline{\rm MS}$-like approximation scheme, in which we treat the ``light'' fermions as (approximately) massless and the ``heavy'' fermions as infinitely heavy. We then obtain 
\begin{equation}
\begin{aligned}
   C_{gg}^{\rm eff}
   &\approx c_{GG} + \frac12\spac\sum_q\spac c_{qq}(m_a)\,\theta(m_a-m_q) 
    = \tilde c_{GG}(m_a) \,, \\
   C_{\gamma\gamma}^{\rm eff}
   &\approx c_{\gamma\gamma} + \sum_f\spac N_c^f\spac Q_f^2\,c_{ff}(m_a)\,\theta(m_a-m_f) 
    = \tilde c_{\gamma\gamma}(m_a) \,,
\end{aligned}
\end{equation}
where the effective couplings on the right-hand side are precisely those defined in (\ref{eq:60}).

\section{Matching onto the chiral Lagrangian}
\label{sec:chiPT}

Using the results derived in the previous sections, the effective ALP Lagrangian (\ref{LlowE}) can be evolved down to scales far below the scale of electroweak symmetry breaking. When one reaches energies of order 1--2\,GeV, only the three light quark flavors $u,d,s$ remain as active degrees of freedom. In order to study the low-energy interactions of a light ALP with hadrons, one should match this Lagrangian onto a chiral effective Lagrangian incorporating the ALP couplings to the light pseudoscalar mesons $(\pi, K, \eta)$. In order to find the bosonized form of the ALP--gluon interaction, one eliminates the $a\spac G\spac \tilde G$ term in favor of ALP couplings to quark bilinears, whose chiral representation is well known. To this end, one performs the chiral rotation \cite{Georgi:1986df,Srednicki:1985xd,Krauss:1986bq,Bardeen:1986yb} 
\begin{equation}\label{chiralrot}
   q\to \exp\bigg(\! -i\bm{\kappa}_q\,c_{GG}\,\frac{a}{f}\,\gamma_5 \bigg)\,q \,, 
\end{equation}
where $q$ is a 3-component vector in flavor space, $\bm{\kappa}_q$ is a real matrix, which we choose to be diagonal in the quark mass basis. Under the chiral rotation the measure of the path integral is not invariant \cite{Fujikawa:1979ay,Fujikawa:1980eg}, and this generates extra terms adding to the anomalous ALP couplings to gluons and photons. Imposing the condition 
\begin{equation}\label{trace}
   \mbox{Tr}\,\bm{\kappa}_q = \kappa_u + \kappa_d + \kappa_s = 1
\end{equation}
ensures that the ALP coupling to $G\spac\tilde G$ is eliminated from the Lagrangian, at the expense of modifying the ALP--photon and ALP--fermion couplings as well as the quark mass matrix. At a scale $\mu_\chi\sim 1$--2\,GeV, this leads to the effective Lagrangian 
\begin{equation}\label{eq:67}
\begin{aligned}
   {\cal L}_{\rm eff}(\mu_\chi)
   &= \frac12 \left( \partial_\mu a\right)\!\left( \partial^\mu a\right) - \frac{m_{a,0}^2}{2}\,a^2
    + \bar q \left( i\rlap{\,/}D - e^{-i\bm{\kappa}_q\,c_{GG}\spac\frac{a}{f}\spac\gamma_5}\spac\bm{m}_q\,
    e^{- i\bm{\kappa}_q\,c_{GG}\spac\frac{a}{f}\spac\gamma_5} \right) q \\
   &\quad\mbox{}+ \frac{\partial^\mu a}{2f}\,\bar q\,\hat{\bm{c}}_{qq} \gamma_\mu\gamma_5\,q 
    + \hat c_{\gamma\gamma}\,\frac{\alpha}{4\pi}\,\frac{a}{f}\,F_{\mu\nu}\,\tilde F^{\mu\nu} + \dots \,,
\end{aligned}
\end{equation}
where $\bm{m}_q=\mbox{diag}(m_u,m_d,m_s)$, and the dots represent the ALP--lepton couplings and possible flavor-changing ALP interactions, both of which are irrelevant to this discussion. The quantities
\begin{equation}\label{hatcpars}
\begin{aligned}
   \hat{\bm{c}}_{qq}(\mu_\chi) & = \bm{c}_{qq}(\mu_\chi) + 2c_{GG}\,\bm{\kappa}_q \,, \\
   \hat{c}_{\gamma\gamma} &= c_{\gamma\gamma} - 2N_c\,c_{GG}\,\sum_q\,Q_q^2\,\kappa_q \,,
\end{aligned}
\end{equation}
with $q=u,d,s$, are the modified ALP--fermion and ALP--photon couplings, whose explicit expressions in terms of the ALP couplings at the UV scale can directly be obtained from the results compiled in the previous sections. The effective Lagrangian (\ref{eq:67}) is equivalent to the original Lagrangian (\ref{LlowE}) evolved to the low scale $\mu_\chi$ and it describes the same physics, even though the ALP coupling to gluons has been removed at the Lagrangian level. The ALP interactions with quarks now enter in two places: in the derivative couplings proportional to the parameters $\hat c_{qq}$, and through the phase factors multiplying the quark mass matrix. Note that the choice of the $\kappa_q$ parameters is completely arbitrary as long as the constraint (\ref{trace}) is satisfied. Below we will demonstrate with two examples that the results obtained for physical observables are indeed independent of the $\kappa_q$ parameters.

As a side remark, let us mention that the effective ALP--pion Lagrangian can also be derived starting from the alternative form of the effective Lagrangian shown in (\ref{Leffalt}). This Lagrangian differs from the original one in (\ref{Leff}) by a chiral rotation of the same form as that shown in (\ref{chiralrot}), but with a different choice of the $\kappa_q$ parameters not subject to the condition (\ref{trace}). A second chiral rotation is then required to eliminate the ALP--gluon coupling. The resulting ALP--pion Lagrangian is equivalent to the one derived here. 

Let us now discuss the matching of the effective Lagrangian (\ref{eq:67}) onto a chiral effective Lagrangian, working consistently at lowest order in the chiral expansion. The Dirac Lagrangian for the quark fields matches onto the standard form of the Gasser--Leutwyler Lagrangian \cite{Gasser:1984gg}, but with the mass matrix replaced by the ALP-field dependent matrix
\begin{equation}\label{hatmq}
   \hat{\bm{m}}_q(a) = \exp\bigg(\! -i\bm{\kappa}_q\,c_{GG}\,\frac{a}{f} \bigg)\,\bm{m}_q\spac 
    \exp\bigg( - i\bm{\kappa}_q\,c_{GG}\,\frac{a}{f} \bigg) \,.
\end{equation}
Next, the axial-vector currents in the derivative couplings of the ALP to quark fields can be matched onto chiral currents using the replacement rules
\begin{equation}
   \bar q_L^{\,i}\spac\gamma_\mu\spac q_L^{\,j} 
   \to - \frac{if_\pi^2}{4} \left[ \bm{\Sigma}\,D_\mu \bm{\Sigma}^\dagger \right]^{ji} , \qquad
   \bar q_R^{\,i}\spac\gamma_\mu\spac q_R^{\,j} 
   \to - \frac{if_\pi^2}{4} \left[ \bm{\Sigma}^\dagger D_\mu\bm{\Sigma} \right]^{ji} .
\end{equation}
In this way, one obtains \cite{Georgi:1986df,Krauss:1986bq,Bardeen:1986yb} 
\begin{equation}\label{LeffChpT}
\begin{aligned}
   {\cal L}_{\chi PT}^{\rm ALP} 
   &= \frac12\,\partial^\mu a\,\partial_\mu a - \frac{m_{a,0}^2}{2}\,a^2
    + \frac{f_\pi^2}{8}\,\mbox{Tr}\big[ D^\mu\bm{\Sigma}\,D_\mu\bm{\Sigma}^\dagger \big] 
    + \frac{f_\pi^2}{4}\,B_0\,\mbox{Tr}\big[ \bm{\Sigma}\,\hat{\bm{m}}_q^\dagger(a) 
    + \hat{\bm{m}}_q(a)\,\bm{\Sigma}^\dagger \big] \\
   &\quad\mbox{}+ \frac{if_\pi^2}{4}\,\frac{\partial^\mu a}{2f}\,
    \mbox{Tr}\big[ \hat{\bm{c}}_{qq} (\bm{\Sigma}\,D_\mu\bm{\Sigma}^\dagger 
    - \bm{\Sigma}^\dagger D_\mu\bm{\Sigma}) \big]
    + \hat c_{\gamma\gamma}\,\frac{\alpha}{4\pi}\,\frac{a}{f}\,F_{\mu\nu}\,\tilde F^{\mu\nu} + \dots ,
\end{aligned}
\end{equation} 
where $\bm{\Sigma}(x)=\exp\big(\frac{i\sqrt2}{f_\pi}\,\lambda_a\spac\pi_a(x)\big)$ contains the pseudoscalar meson fields ($\lambda_a$ are the Gell-Mann matrices), and the parameter $B_0\approx m_\pi^2/(m_u+m_d)$ is proportional to the chiral condensate. The covariant derivative is defined as $D_\mu\bm{\Sigma}=\partial_\mu\bm{\Sigma}-ieA_\mu [\bm{Q},\bm{\Sigma}]$, where $\bm{Q}=\text{diag}(Q_u,Q_d,Q_s)$. 

For the case of the QCD axion (with $m_{a,0}^2=0$), the chiral effective ALP Lagrangian was first introduced in \cite{Georgi:1986df} and has recently been studied in great detail in \cite{diCortona:2015ldu}. In general, the last term in the first line of (\ref{LeffChpT}) gives rise to a mass mixing of the ALP with the pseudoscalar mesons $\pi^0$ and $\eta_8$. In order to eliminate this mixing, one chooses the matrix $\bm{\kappa}_q$ in such a way that $\bm{\kappa}_q\spac\bm{m}_q\propto\mathbbm{1}$. When combined with the condition (\ref{trace}) this implies
\begin{equation}\label{kappaqchoice}
\begin{aligned}
   \kappa_u &= \frac{m_d\spac m_s}{m_u\spac m_d+m_u\spac m_s+m_d\spac m_s} 
    \approx \frac{m_d}{m_u+m_d} \,, \\
   \kappa_d &= \frac{m_u\spac m_s}{m_u\spac m_d+m_u\spac m_s+m_d\spac m_s}
    \approx \frac{m_u}{m_u+m_d} \,, \\
   \kappa_s &= \frac{m_u\spac m_d}{m_u\spac m_d+m_u\spac m_s+m_d\spac m_s}
    \approx \frac{m_u m_d}{(m_u+m_d)\,m_s} \ll \kappa_{u,d} \,.
\end{aligned}
\end{equation}
With this choice, the modified ALP--photon coupling takes the form
\begin{equation}\label{E/N}
   \hat c_{\gamma\gamma} = c_{\gamma\gamma}
    - \frac23\,\frac{4m_d\spac m_s+m_u(m_s+m_d)}{m_u\spac m_d+m_u\spac m_s+m_d\spac m_s}\,c_{GG}  
   \simeq c_{\gamma\gamma} - 2.0\,c_{GG} \,,
\end{equation}
where we have used the ratios $m_u/m_d=0.49\pm 0.02$ and $2m_s/(m_u+m_d)=27.4\pm 0.1$ \cite{Zyla:2020zbs}. Next-to-leading order corrections in the chiral expansion to $\hat c_{\gamma\gamma}$ have been worked out in \cite{diCortona:2015ldu}. They reduce the coefficient in front of $c_{GG}$ to $-(1.92\pm 0.04)$. At lowest order in the chiral expansion one finds that, with the above choice of $\kappa_q$ values, there are no additional contributions to the $a\to\gamma\gamma$ decay amplitude beyond those governed by $\hat c_{\gamma\gamma}$ in (\ref{E/N}). QCD dynamics generates a mass for the ALP, thereby breaking the continuous shift symmetry of the classical Lagrangian. Expanding the terms in the first line of (\ref{LeffChpT}) to quadratic order in the pion and ALP fields, one finds \cite{Bardeen:1978nq,Shifman:1979if,DiVecchia:1980yfw}
\begin{equation}
   m_a^2 = c_{GG}^2\,\frac{f_\pi^2\,m_\pi^2}{f^2}\,\frac{2m_u m_d}{(m_u+m_d)^2} 
   = \frac{f_\pi^2\,m_\pi^2}{2f_a^2}\,\frac{m_u m_d}{(m_u+m_d)^2} \,,
\end{equation} 
up to higher-order corrections in the chiral expansion and independent of the choice of the individual $\kappa_q$ values. Here $f_a=-f/(2c_{GG})$ is the axion decay constant. More generally, one finds that the axion potential is a periodic function of the axion field, which is invariant under the discrete shift transformation $a\to a+2n\pi f_a$.

\begin{figure}
\begin{center}
\includegraphics[height=2.2cm]{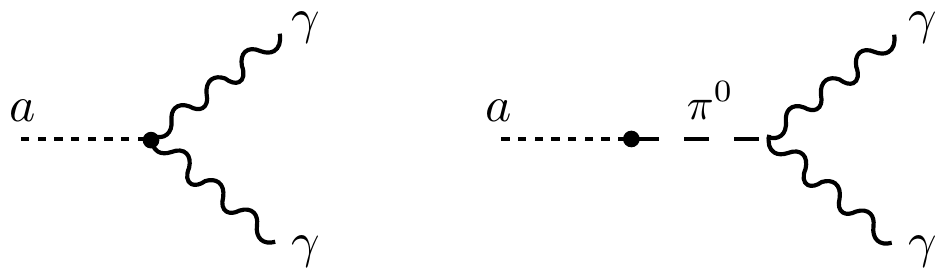}
\vspace{2mm}
\caption{\label{fig:chiPT} 
Leading-order contributions to the $a\to\gamma\gamma$ decay amplitude in the chiral expansion. The $\pi^0\gamma\gamma$ coupling is obtained from the Wess--Zumino--Witten term not shown explicitly in (\ref{LeffChpT}).} 
\end{center}
\end{figure}

Let us now discuss the structure of the effective chiral Lagrangian for a light ALP in the presence of a non-vanishing mass term $m_{a,0}^2$. As we will show, in this case some non-trivial complications arise from the mixing of the ALP with the pseudoscalar mesons, which give rise to an additional contribution to the $a\to\gamma\gamma$ decay amplitude \cite{Bauer:2017ris}. For simplicity, we will consider the case of two light flavors $u$ and $d$. The generalization to three flavors is straightforward, but the additional contributions one finds are suppressed by $m_{u,d}/m_s$. In our discussion we {\em do not\/} impose the relations $\kappa_u=m_d/(m_u+m_d)$ and $\kappa_d=m_u/(m_u+m_d)$, which would be the analogue of (\ref{kappaqchoice}) for the case of two flavors. Physical quantities are independent of the choice of the $\kappa_q$ parameters, and it is instructive to trace in detail how the dependence on these parameters, which enters via the matrix $\hat{\bm{m}}_q(a)$ and via the coupling parameters $\hat{\bm{c}}_{qq}$ and $\hat c_{\gamma\gamma}$ in (\ref{hatcpars}), cancels out. As an important example, we consider the decay $a\to\gamma\gamma$. As shown in Figure~\ref{fig:chiPT}, at leading order in chiral perturbation theory there are two contributions to the decay amplitude: one involving the coupling $\hat c_{\gamma\gamma}$ and one involving the mixing of the ALP with the neutral pion, followed by the decay $\pi^0\to\gamma\gamma$ mediated by the axial anomaly. The latter coupling can be implemented in the chiral Lagrangian through the Wess--Zumino--Witten term \cite{Witten:1983tx}. Combining the two contributions, we find
\begin{equation}
\begin{aligned}
   {\cal M}(a\to\gamma\gamma)
   &= \langle\gamma\gamma|\,F_{\mu\nu}\,\tilde F^{\mu\nu}\,|0\rangle\,
    \bigg[ \frac{\alpha}{4\pi}\,\frac{\hat c_{\gamma\gamma}}{f} \\
   &\quad\mbox{}+ \frac{f_\pi}{\sqrt2\spac f} \left( \frac{\hat c_{uu}-\hat c_{dd}}{2}\,p_a^2
    - 2c_{GG}\,\frac{m_u\spac\kappa_u-m_d\spac\kappa_d}{m_u+m_d}\,m_\pi^2 \right) 
    \frac{i}{p_a^2-m_\pi^2}\,\bigg(\! - \frac{i\sqrt2}{f_\pi}\,\frac{\alpha}{4\pi} \bigg) \bigg] \,,
\end{aligned}
\end{equation} 
where $p_a^2=m_a^2$, and the last factor in the second line arises from the $\pi^0\to\gamma\gamma$ vertex.\footnote{Note that the result for this vertex given in eq.\,(19.117) of \cite{Peskin:1995ev} misses a minus sign, see the errata pages listed in this reference.} 
Adding up the various terms inside the bracket in the first line we find that any dependence on the parameters $\kappa_q$ cancels out, and one is left with the combination 
\begin{equation}\label{eq:76}
   C_{\gamma\gamma}^{\rm eff} = c_{\gamma\gamma} 
    - \left( \frac53 + \frac{m_\pi^2}{m_\pi^2-m_a^2}\,\frac{m_d-m_u}{m_u+m_d} \right) c_{GG}
    - \frac{m_a^2}{m_\pi^2-m_a^2}\,\frac{c_{uu}(\mu_\chi)-c_{dd}(\mu_\chi)}{2} \,,
\end{equation}
which was identified in \cite{Bauer:2017ris} as the effective ALP coupling to photons in the chiral effective theory. Note that there are additional contributions from the charged leptons, which have been given in (\ref{rates}) but are not included here. In the limit of a very light ALP ($m_a^2\ll m_\pi^2$) the above relation reduces to (\ref{E/N}) in the approximation where the strange-quark mass is decoupled. For a heavier ALP, however, the additional contributions can be important. For $m_a^2\gg m_\pi^2$ we obtain $C_{\gamma\gamma}^{\rm eff}\simeq c_{\gamma\gamma}-1.67 c_{GG}+0.5\spac(c_{uu}-c_{dd})$, which is now explicitly dependent on the quark couplings. The contribution proportional to the mass difference of the up and down quarks in (\ref{eq:76}) results from the coupling of the neutral pion to $G_{\mu\nu}^a\,\tilde G^{\mu\nu,a}$. The corresponding matrix element can be derived using the anomaly equation and assuming isospin invariance of the pion matrix elements of axial-vector and pseusoscalar currents \cite{Beneke:2000ry,Beneke:2002jn}. One finds
\begin{equation}
   \big\langle\pi^0\big|\,\frac{\alpha_s}{4\pi}\,\frac{a}{f}\,G_{\mu\nu}^a\,\tilde G^{\mu\nu,a}\,
    \big|a\big\rangle
   = \frac{1}{f}\,\big\langle\pi^0\big|\,\frac{\alpha_s}{4\pi}\,G_{\mu\nu}^A\,\tilde G^{\mu\nu,A}\,
    \big|0\big\rangle
   = \frac{1}{f}\,\frac{m_d-m_u}{m_d+m_u}\,\frac{f_\pi\,m_\pi^2}{\sqrt2} \,.
\end{equation}
The pion then decays into two photons via the axial anomaly. The contribution 5/3 in (\ref{eq:76}) arises from an analogous coupling to the flavor-singlet meson $\varphi^0$ (the analogue of $\eta_1$ in flavor $SU(3)$) \cite{Kaiser:2000gs}. The result (\ref{eq:76}) implicitly assumes that the ALP is lighter than the flavor-singlet mesons, because these have been integrated out from the chiral Lagrangian. In the opposite limit one should use the perturbative expression shown in (\ref{eq:47a}) for the effective ALP--photon coupling. Finally, the contribution proportional to the $c_{qq}$ parameters is due to the kinetic mixing of the ALP with the neutral pion. This effect introduces a dependence of the effective ALP--photon coupling on the parameters $c_{qq}$, which is absent for the QCD axion. Note that the difference of the ALP couplings to up and down quarks receives an important contribution from RG evolution. From (\ref{boundlessbeauty}) we find the approximate expression
\begin{equation}
   c_{uu}(\mu_\chi) - c_{dd}(\mu_\chi) 
   \approx c_{uu}(\Lambda) - c_{dd}(\Lambda) 
    - 6\,\frac{\alpha_t(m_t)}{\alpha_s(m_t)}\,
    \bigg[ 1 - \left( \frac{\alpha_s(\Lambda)}{\alpha_s(m_t)} \right)^{\!\frac17} \bigg]\,
    c_{tt}(\Lambda) \,.
\end{equation}

In order to avoid the presence of ALP--pion mixing contributions in perturbative calculations, one needs to diagonalize the quadratic terms in the effective chiral Lagrangian. Upon expanding the Lagrangian to quadratic order in fields, we obtain 
\begin{equation}
   {\cal L}_{\chi PT}^{\rm ALP} 
   \ni \frac12\,(\partial^\mu\varphi)^T \bm{Z}\,(\partial_\mu\varphi)
    - \frac12\,\varphi^T\hspace{-0.3mm}\bm{M}^2\spac\varphi \,; \quad \text{with} \quad
   \varphi = \left( \begin{array}{c} \pi^0 \\ a \end{array} \right) ,
\end{equation} 
where the symmetric matrices accounting for kinetic and mass mixing are given by 
\begin{equation}
\begin{aligned}
   \bm{Z} &= \left( \begin{array}{cc} 
    1 & \frac{f_\pi}{\sqrt 2 f}\spac\frac{\hat c_{uu}-\hat c_{dd}}{2} \\ 
    \frac{f_\pi}{\sqrt 2 f}\spac\frac{\hat c_{uu}-\hat c_{dd}}{2} & 1 \end{array} \right) , \\
   \bm{M}^2 &= m_{\pi,0}^2 \left( \begin{array}{ccc} 
    1 && \frac{\sqrt 2 f_\pi\spac c_{GG}}{f}\spac\frac{m_u\spac\kappa_u-m_d\spac\kappa_d}{m_u+m_d} \\ 
    \frac{\sqrt 2 f_\pi\spac c_{GG}}{f}\spac\frac{m_u\spac\kappa_u-m_d\spac\kappa_d}{m_u+m_d} &~&
     \frac{m_{a,0}^2}{m_{\pi,0}^2} 
     + \frac{2f_\pi^2\spac c_{GG}^2}{f^2}\spac\frac{m_u\spac\kappa_u^2+m_d\spac\kappa_d^2}{m_u+m_d}
    \end{array} \right) .
\end{aligned}
\end{equation} 
The parameter $m_{\pi,0}^2=B_0\spac(m_d+m_u)$ gives the leading-order contribution to the pion mass. In order to find the properly normalized mass eigenstates, we first diagonalize the matrix $\bm{Z}$, i.e.\ we construct the unitary matrix $\bm{U}_Z$ such that $\bm{U}_Z^\dagger\spac\bm{Z}\,\bm{U}_Z=\bm{Z}_{\rm diag}$. We then rescale the fields to bring the kinetic terms into a canonical form. In the final step we diagonalize the resulting mass matrix $\hat{\bm{M}}^2\equiv\bm{Z}_{\rm diag}^{-1/2}\,\bm{U}_Z^\dagger\spac\bm{M}^2\spac\bm{U}_Z\spac\bm{Z}_{\rm diag}^{-1/2}$, i.e.\ we construct the unitary matrix $\bm{U}_M$ such that $\bm{U}_M^\dagger\spac\hat{\bm{M}}^2\spac\bm{U}_M=\bm{M}_{\rm diag}^2$. The physical mass eigenstates are related to the original fields in the chiral Lagrangian by $\varphi_{\rm phys}=\bm{U}_M^\dagger\spac\bm{Z}_{\rm diag}^{1/2}\,\bm{U}_Z^\dagger\,\varphi$. Written out in components, this leads to\footnote{These relations and (\ref{eq:84}) holds as long as $|m_{\pi,0}^2-m_{a,0}^2|\gg m_{\pi,0}^2\spac f_\pi/f$. In the opposite limit one would obtain maximal mixing, i.e.\ $\pi_{\rm phys}^0=\frac{1}{\sqrt2}\,(\pi^0\pm a)+{\cal O}(f_\pi/f)$. Besides the fact that such a large mixing would require a fine-tuning of the mass parameters that is rather implausible, it would modify the properties of the neutral pion in a way that is incompatible with experimental findings.} 
\begin{equation}\label{physstates}
\begin{aligned}
   \pi_{\rm phys}^0 
   &= \pi^0 - \frac{f_\pi}{2\sqrt2 f} \left( \frac{m_{a,0}^2}{m_{\pi,0}^2-m_{a,0}^2}\,\Delta c_{ud}
    - \delta_\kappa \right) a + {\cal O}\bigg(\frac{f_\pi^2}{f^2}\bigg) \,, \\
   a_{\rm phys}
   &= a + \frac{f_\pi}{2\sqrt2 f}\,\frac{m_{\pi,0}^2}{m_{\pi,0}^2-m_{a,0}^2}\,
    \Delta c_{ud}\spac\,\pi^0 + {\cal O}\bigg(\frac{f_\pi^2}{f^2}\bigg) \,,
\end{aligned}
\end{equation} 
where we have defined
\begin{equation}\label{Deltacud}
   \Delta c_{ud} = c_{uu}(\mu_\chi) - c_{dd}(\mu_\chi) + 2 c_{GG}\,\frac{m_d-m_u}{m_d+m_u} \,, \qquad
   \delta_\kappa = 4 c_{GG}\,\frac{m_u\spac\kappa_u-m_d\spac\kappa_d}{m_d+m_u} \,.
\end{equation}
Importantly, some scheme-dependent terms enter the admixture of an ALP component in the physical pion state. Inverting the first relation, and eliminating the bare mass terms in favor of the physical ones, we find \begin{equation}\label{admixture}
   \pi^0 = \pi_{\rm phys}^0 + \frac{f_\pi}{2\sqrt2\spac f} \left(
    \frac{m_a^2}{m_\pi^2-m_a^2}\,\Delta c_{ud} - \delta_\kappa \right) a_{\rm phys} 
    + {\cal O}\bigg(\frac{f_\pi^2}{f^2}\bigg) \,.
\end{equation} 
The physical masses squared of the neutral pion and the ALP are given by the eigenvalues $m^2$ of the equation $\det(\bm{M}^2-m^2\spac\bm{Z})=0$. We find
\begin{equation}\label{eq:84}
\begin{aligned}
   m_\pi^2 &= m_{\pi,0}^2 \left[ 1 
    + \frac{m_{\pi,0}^2}{m_{\pi,0}^2-m_{a,0}^2}\,\frac{f_\pi^2}{8f^2} 
    \left( \Delta c_{ud} \right)^2 \right] + {\cal O}\bigg(\frac{f_\pi^4}{f^4}\bigg) \,, \\
   m_a^2 &= m_{a,0}^2 \left\{ 1 + \frac{f_\pi^2}{8f^2} \left[
    \left( \Delta c_{ud} + \delta_\kappa \right)^2 
    - \frac{m_{\pi,0}^2}{m_{\pi,0}^2-m_{a,0}^2} \left( \Delta c_{ud} \right)^2 \right] \right\} \\
   &\quad\mbox{}+ m_{\pi,0}^2\,c_{GG}^2\,\frac{f_\pi^2}{f^2}\,\frac{2m_u\spac m_d}{(m_u+m_d)^2} 
    + {\cal O}\bigg(\frac{f_\pi^4}{f^4}\bigg) \,.
\end{aligned}
\end{equation}
In the limit where $m_{a,0}^2\ll m_{\pi,0}^2$ we recover relation (\ref{massrela}). Note that the ALP mass receives a scheme-dependent contribution involving the $\kappa_q$ parameters. This is not a problem, because only the physical mass parameter $m_a^2$ is observable, whereas the ``bare'' mass parameter $m_{a,0}^2$ is not. 

For the special choice $\kappa_u=m_d/(m_u+m_d)$ and $\kappa_d=m_u/(m_u+m_d)$, the quantity $\delta_\kappa$ vanishes and relation (\ref{admixture}) reduces to a result derived in \cite{Bauer:2017ris}. But this choice does not eliminate the ALP--pion mixing. Instead, in the presence of a non-vanishing ALP mass the optimal choice of the $\kappa_q$ parameters is
\begin{equation}\label{preferred}
   \kappa_u = \frac{m_d}{m_u+m_d} + \frac{m_a^2}{m_\pi^2-m_a^2}\,\frac{\Delta c_{ud}}{4c_{GG}} \,, \qquad
   \kappa_d = \frac{m_u}{m_u+m_d} - \frac{m_a^2}{m_\pi^2-m_a^2}\,\frac{\Delta c_{ud}}{4c_{GG}} \,. 
\end{equation} 
In the limit where $m_a^2/m_\pi^2\to 0$ this reduces to the default choice usually adopted in the literature, but for generic ALP masses the additional contributions introduce important corrections. With the choice (\ref{preferred}) the physical neutral-pion state does not contain an admixture of the ALP at first order in $f_\pi/f$, the parameter $\hat c_{\gamma\gamma}$ in the effective Lagrangian (\ref{LeffChpT}) agrees with the effective ALP--photon coupling $C_{\gamma\gamma}^{\rm eff}$ shown in (\ref{eq:76}), and the parameters $\hat c_{qq}$ satisfy the relation
\begin{equation}
   \hat c_{uu} - \hat c_{dd}
   = \frac{m_\pi^2}{m_\pi^2-m_a^2}\,\Delta c_{ud} \,.
\end{equation}
Finally, with this choice the physical ALP mass can be expressed as 
\begin{equation}
   m_a^2 = m_{a,0}^2 \left[ 1 + \frac{f_\pi^2}{8f^2}\, 
    \frac{m_\pi^2\,m_{a,0}^2}{\left(m_\pi^2-m_{a,0}^2\right)^2} \left( \Delta c_{ud} \right)^2 \right] 
    + c_{GG}^2\,\frac{f_\pi^2\,m_\pi^2}{f^2}\,\frac{2m_u\spac m_d}{(m_u+m_d)^2} 
   + {\cal O}\bigg(\frac{f_\pi^4}{f^4}\bigg) \,.
\end{equation} 
This result generalizes relation (\ref{massrela}) to arbitrary values of the Lagrangian parameter $m_{a,0}^2$.

When the effective chiral Lagrangian (\ref{LeffChpT}) is expressed in terms of the physical states given in (\ref{physstates}), one finds (now for general $\kappa_q$ parameters) 
\begin{align}\label{LChPToptimal}
   {\cal L}_{\chi PT}^{\rm ALP} 
   &= \frac12\,\partial^\mu a\,\partial_\mu a - \frac{m_a^2}{2}\,a^2
    + \frac12\,\partial^\mu\pi^0\spac\partial_\mu\pi^0 - \frac{m_\pi^2}{2}\,\big( \pi^0 \big)^2
    + D^\mu\pi^+\spac D_\mu\pi^- - m_\pi^2\,\pi^+\pi^- 
    + {\cal O}\!\left( \frac{\pi^4}{f_\pi^2} \right) \notag\\
   &\quad\mbox{}+ \frac{\Delta c_{ud}}{6\sqrt 2 f_\pi f}\,\frac{1}{m_a^2-m_\pi^2}\,\Big[\spac
    2\spac(m_a^2-2m_\pi^2)\,\partial^\mu a \left( \pi^0\pi^+ D_\mu\pi^-
    + \pi^0\pi^- D_\mu\pi^+ - 2\pi^+\pi^- \partial_\mu\pi^0 \right) \notag\\[-0.5mm]
   &\hspace{4.35cm}\mbox{}- 2 m_a^2\,a \left( \pi^+ D^\mu\pi^-\partial_\mu\pi^0 
    + \pi^- D^\mu\pi^+\partial_\mu\pi^0 - 2\pi^0 D^\mu\pi^+ D_\mu\pi^- \right) \notag\\
   &\hspace{4.35cm}\mbox{}- m_a^2\spac m_\pi^2\,a \left( 2\pi^+\pi^-\pi^0 
    + \left( \pi^0 \right)^3 \right) \Big] + {\cal O}\!\left( \frac{a\spac\pi^5}{f_\pi^3 f} \right) 
    \notag\\[-1mm]
   &\quad\mbox{}+ \frac{\delta_\kappa}{3\sqrt 2 f_\pi f}\,\Big[
    \partial^\mu a \left( \pi^0\pi^+ D_\mu\pi^- + \pi^0\pi^- D_\mu\pi^+ 
    - 2\pi^+\pi^- \partial_\mu\pi^0 \right) \notag\\[-2mm]
   &\hspace{2.65cm}\mbox{}- a \left( \pi^+ D^\mu\pi^-\partial_\mu\pi^0 
    + \pi^- D^\mu\pi^+\partial_\mu\pi^0 - 2\pi^0 D^\mu\pi^+ D_\mu\pi^- \right) \Big] \notag\\
   &\quad\mbox{}+ \left( \hat c_{\gamma\gamma} - \frac{\delta_\kappa}{2} 
    + \frac{m_a^2}{m_a^2-m_\pi^2}\,\frac{\Delta c_{ud}}{2} \right)
    \frac{\alpha}{4\pi}\,\frac{a}{f}\,F_{\mu\nu}\,\tilde F^{\mu\nu} 
    + \dots \,,
\end{align} 
where $D_\mu\pi^\pm=(\partial_\mu\mp ie A_\mu)\pi^\pm$, and for simplicity we have suppressed the subscript ``phys'' on the fields. The coefficient in front of the ALP--photon coupling, which is the sum of the coefficient $\hat c_{\gamma\gamma}$ and a contribution from the Wess--Zumino--Witten term, is nothing but the quantity $C_{\gamma\gamma}^{\rm eff}$ given in (\ref{eq:76}). It is independent of the choice of the $\kappa_q$ parameters. The remaining dependence, which enters via the quantity $\delta_\kappa$, drops out when one calculates physical matrix elements. Indeed, using integration by parts it can be seen that the coefficient of $\delta_\kappa$ vanishes when the equations of motion for the pion fields are used. They can thus be dropped from the effective Lagrangian. It follows that a single parameter $\Delta c_{ud}$ governs the leading-order interactions of the ALP with pions, and we obtain the final expression
\begin{align}
   {\cal L}_{\chi PT}^{\rm ALP} 
   &= \frac12\,\partial^\mu a\,\partial_\mu a - \frac{m_a^2}{2}\,a^2
    + \frac12\,\partial^\mu\pi^0\spac\partial_\mu\pi^0 - \frac{m_\pi^2}{2}\,\big( \pi^0 \big)^2
    + D^\mu\pi^+\spac D_\mu\pi^- - m_\pi^2\,\pi^+\pi^- 
    + {\cal O}\!\left( \frac{\pi^4}{f_\pi^2} \right) \notag\\
   &\quad\mbox{}- \frac{\Delta c_{ud}}{6\sqrt 2 f_\pi f}\,\frac{m_\pi^2}{m_a^2-m_\pi^2}\,\Big[\spac 
    4\spac\partial^\mu a \left( \pi^0\pi^+ D_\mu\pi^- + \pi^0\pi^- D_\mu\pi^+ 
    - 2\pi^+\pi^- \partial_\mu\pi^0 \right) \notag\\[-1mm]
   &\hspace{4.35cm}\mbox{}+ m_a^2\,a \left( 2\pi^+\pi^-\pi^0 
    + \left( \pi^0 \right)^3 \right) \Big] + {\cal O}\!\left( \frac{a\spac\pi^5}{f_\pi^3 f} \right) 
    \notag\\[-1mm]
   &\quad\mbox{}+ C_{\gamma\gamma}^{\rm eff}\,\frac{a}{f}\,F_{\mu\nu}\,\tilde F^{\mu\nu} 
    + \dots \,.
\end{align} 
This generalizes the effective axion--pion Lagrangian derived in \cite{Chang:1993gm} to the case of an ALP with non-zero mass parameter $m_a^2$.

\begin{figure}
\begin{center}
\includegraphics[height=6.5cm]{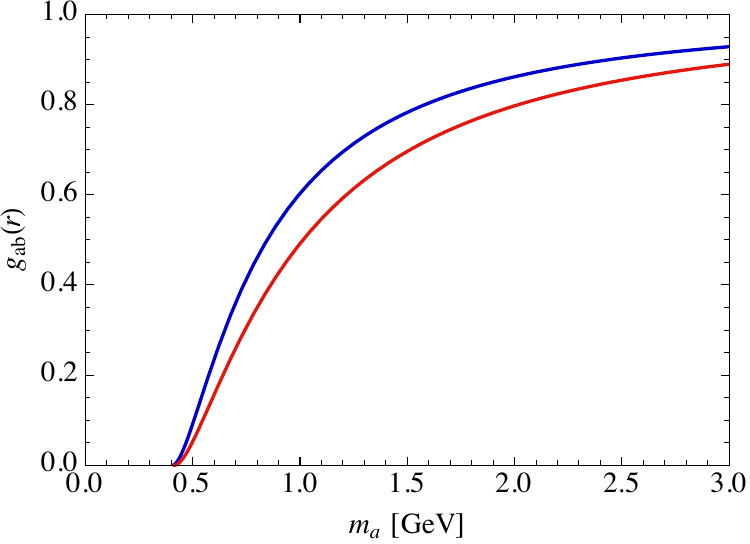}
\caption{\label{fig:a3pi} 
Dependence of the phase-space functions $g_{00}(r)$ (blue) and $g_{+-}(r)$ (red) on the ALP mass (with $r=m_\pi^2/m_a^2$).} 
\end{center}
\end{figure}

As an important application of the Lagrangian (\ref{LChPToptimal}) we consider the decays of an ALP into three pions, which is allowed if the ALP mass is larger than $3m_\pi$. We obtain the decay amplitudes
\begin{equation}
\begin{aligned}
   {\cal M}(a\to\pi^0\pi^0\pi^0)
   &= - \frac{\Delta c_{ud}}{\sqrt 2 f_\pi f}\,\frac{m_\pi^2\spac m_a^2}{m_a^2-m_\pi^2} \,, \\
   {\cal M}(a\to\pi^+\pi^-\pi^0)
   &= - \frac{\Delta c_{ud}}{\sqrt 2 f_\pi f}\,\frac{m_\pi^2\spac(m_{+-}^2-m_\pi^2)}{m_a^2-m_\pi^2} \,,
\end{aligned}
\end{equation} 
where $m_{+-}^2=(p_{\pi^+}+p_{\pi^-})^2$ is the invariant mass squared of the charged pion pair. These expressions agree with corresponding results derived in \cite{Bauer:2017ris}. In this reference also the differential distributions in the Dalitz plot were derived. Note that the chiral expansion makes sense only in the region of phase space where the pion momenta are small compared with the scale of chiral symmetry breaking, $4\pi f_\pi\simeq 1.63$\,GeV. This requires the ALP to be lighter than about 3\,GeV. For the total decay rates one finds 
\begin{equation}
   \Gamma(a\to\pi^a\pi^b\pi^0) 
   = \frac{m_a\spac m_\pi^4}{6144\pi^3 f_\pi^2\spac f^2} \left( \Delta c_{ud} \right)^2
    g_{ab}\bigg(\frac{m_\pi^2}{m_a^2}\bigg) \,,
\end{equation}
where (with $0\le r\le 1/9$)
\begin{equation}
\begin{aligned}
   g_{00}(r) &= \frac{2}{(1-r)^2} \int_{4r}^{(1-\sqrt{r})^2}\!\!dz\,\sqrt{1-\frac{4r}{z}}\,
    \lambda^{1/2}(1,z,r) \,, \\
   g_{+-}(r) &= \frac{12}{(1-r)^2} \int_{4r}^{(1-\sqrt{r})^2}\!\!dz\,\sqrt{1-\frac{4r}{z}}\,(z-r)^2\,
    \lambda^{1/2}(1,z,r) \,.
\end{aligned}
\end{equation}
Both functions are normalized such that $g_{ab}(0)=1$, and they vanish at the threshold $r=1/9$. The dependence of these two functions on the ALP mass is shown in Figure~\ref{fig:a3pi}. Interestingly, the two decay rates are almost of equal size, despite of the fact that the rate of the $a\to 3\pi^0$ mode contains a symmetry factor $1/6$. From a phenomenological point of view the $a\to 3\pi$ decay rates can be important. For $m_a=1$\,GeV, we find that $\Gamma(a\to 3\pi)/\Gamma(a\to\gamma\gamma)\simeq 4.6\,(\Delta c_{ud}/C_{\gamma\gamma}^{\rm eff})^2$, where the ratio of couplings is naturally of ${\cal O}(1)$, see (\ref{eq:76}) and (\ref{Deltacud}).

\section{Conclusions}

Axions and axion-like particles (commonly referred to as ALPs in this work) are well-motivated new-physics candidates in extensions of the Standard Model (SM) with a spontaneously broken global symmetry. In these models the mass scale of the new-physics sector is set by the scale at which the global symmetry is broken, whereas the mass of the associated pseudo Nambu--Goldstone boson (the ALP) can be significantly smaller. The fundamental coupling structure of an ALP is therefore determined at the ultra-violet (UV) scale $\Lambda$, while most experimental searches are performed at energies comparable to its mass. The couplings at this low scale dictate the most relevant interactions of an ALP, its branching ratios and the most promising search strategies. In this work we have derived the low-energy ALP couplings by starting from the most general Lagrangian including all leading-order dimension-5 operators at the UV scale, systematically evolving the coupling parameters to lower energies, and matching onto an effective Lagrangian below the electroweak scale and finally to the chiral Lagrangian. The corresponding equations represent a complete framework for calculating the couplings of an ALP to SM particles at any given scale. 
 
At the UV scale the effective ALP Lagrangian can be defined in terms of different but equivalent operator bases, which make manifest either the derivative nature of the ALP couplings or the suppression of the ALP--fermion interactions by the fermion masses. We have demonstrated the equivalence of the different bases explicitly, using the example of the one-loop ALP decay widths into gauge bosons. The most general effective Lagrangian contains a redundant operator, in which the ALP is derivatively coupled to the Higgs current. This operator can be reduced to the remaining operators using field redefinitions. While it can thus consistently be omitted from the operator basis, the presence of UV divergences in the three-point function connecting an ALP with two scalar fields must nevertheless be treated carefully when deriving the evolution equations for the ALP couplings to fermions.

We have presented the renormalization-group (RG) evolution equations for the ALP couplings above and below the electroweak scale, working consistently at two-loop order in gauge couplings and one-loop order in Yukawa interactions. In our default basis shown in (\ref{Leff}), the ALP--boson couplings are scale independent in this order, while the ALP--fermion couplings obey a rather complicated set of coupled differential equations, given in (\ref{RGEs}). We have derived the exact solution to these equations in the approximation where one neglects the SM Yukawa couplings of all fermions with the exception of the top-quark, for which $y_t\approx 1$. At the electroweak scale, we have expressed the effective Lagrangian in terms of fields in the broken phase, which correspond to the mass eigenstates of the SM particles. High-energy collider processes involving ALPs can be systematically studied using the ALP effective Lagrangian formulated at this scale.

For low-energy processes involving ALPs, it is necessary to evolve the effective Lagrangian down to lower scales. At the electroweak scale we have performed a systematic matching procedure by integrating out the top quark, the Higgs boson and the $W$ and $Z$ bosons. We have included all matching contributions at one-loop order and partially accounted for some two-loop contributions required by RG invariance of the effective theory. A particularly important class of matching contributions gives rise to flavor-changing ALP--fermion couplings induced by $W$-boson exchange, which exists even if the underlying UV theory is flavor universal or, more generally, respects the principle of minimal flavor violation. The RG evolution below the electroweak scale only affects the flavor-diagonal ALP--fermion couplings and arises at order $\alpha_s^2$ and $\alpha^2$ in the QCD and QED running couplings. At the scale of chiral symmetry breaking we have matched the effective Lagrangian onto the chiral Lagrangian extended with an ALP field. We have discussed ALP--pion mixing, the effective ALP--photon coupling of a light ALP (with mass below the GeV scale) and the leading-order ALP couplings to three pions in the presence of a non-zero mass term for the ALP in the UV theory. We have emphasized that in this case the optimal choice of the chiral rotation which eliminates the ALP--gluon coupling as well the ALP mixing with the $\pi^0$ state is different from the case of the classical QCD axion. We have also shown explicitly that the ALP--photon coupling and the ALP couplings to pions are independent of the parameters of the chiral rotation. 

There is an important flavor-universal contribution to the ALP couplings to fermions, which is generated above the scale of the top-quark mass and arises from the ALP mixing into the neutral SM Goldstone boson. This effect is induced through top-quark loops and receives large logarithmic corrections of order $\alpha_t^n\ln^n(\Lambda^2/m_t^2)$ in higher orders of perturbation theory, which we have resummed to all orders, see (\ref{boundlessbeauty}). It generates, for example, a sizable ALP--electron coupling in the low-energy theory  even if only ALP couplings to quark doublets or right-handed up-type quarks are present in the UV theory. Depending on the ALP mass, experimental searches for $a\to e^+ e^-$ decays, astrophysical constraints or precision-spectroscopy searches sensitive to the simultaneous presence of ALP--electron and ALP--nucleon couplings can discover an ALP with these properties even if the ALP does not interact with leptons at the UV scale. This argument extends to the case of the QCD axion, for which we have shown that the DFSZ axion has sizable couplings to electrons even at low $\tan\beta$, where the UV coupling to electrons is strongly suppressed.

The results of this paper form the basis for precise phenomenological analyses of the physics of axions and ALPs, connecting low-energy observables in a systematic and accurate way with the couplings of the underlying UV theory. 

\subsubsection*{Acknowledgments}

M.N.~thanks Gino Isidori, the particle theory group at Zurich University and the Pauli Center for hospitality during a sabbatical stay. The research of M.N.\ and M.S.\ was supported by the Cluster of Excellence {\em Precision Physics, Fundamental Interactions and Structure of Matter\/} (PRISMA${}^+$ -- EXC~2118/1) within the German Excellence Strategy (project ID 39083149). S.R.\ acknowledges support from the INFN grant no.~SESAMO.

\subsubsection*{Note added}
 
During the final stages of this project another study appeared \cite{Chala:2020wvs}, in which the RG evolution of the ALP Lagrangian is discussed at one-loop order. After this paper has appeared on arXiv.org, we learned that the two-loop contributions involving the ALP--fermion couplings, which are included in the evolution equations (\ref{RGEs}) via our definitions of the parameters $\tilde c_{VV}$ in (\ref{rela2}), have also been discussed in \cite{Choi:2020rgn}.

\newpage
\begin{appendix}

\section{\boldmath Scale dependence of the $\tilde c_{VV}$ parameters}
\label{app:A}
\renewcommand{\theequation}{A.\arabic{equation}}
\setcounter{equation}{0}

Contrary to the original ALP--boson couplings $c_{VV}$, the quantities $\tilde c_{VV}$ defined in (\ref{rela2}) are no longer scale independent at two-loop order, but they satisfy the evolution equations \begin{align}\label{RGEs2l}
   \frac{d\spac\tilde c_{GG}}{d\ln\mu}
   &= \sum_q\spac\frac{y_q^2}{8\pi^2}\,c_{qq}
    + \frac{9\alpha_s^2}{2\pi^2}\,C_F^{(3)}\spac\tilde c_{GG}
    + \frac{9\alpha_2^2}{4\pi^2}\,C_F^{(2)}\spac\tilde c_{WW}
    + \frac{9\alpha_1^2}{8\pi^2} \left( {\cal Y}_u^2 + {\cal Y}_d^2 
    + 2{\cal Y}_Q^2 \right) \tilde c_{BB} \notag\\
   &\approx \frac{y_t^2}{8\pi^2}\,c_{tt}
    + \frac{6\alpha_s^2}{\pi^2}\,\tilde c_{GG} 
    + \frac{27\alpha_2^2}{16\pi^2}\,\tilde c_{WW}
    + \frac{11\alpha_1^2}{16\pi^2}\,\tilde c_{BB} \,, \notag\\
   \frac{d\spac\tilde c_{WW}}{d\ln\mu}
   &= \sum_q\spac\frac{3y_q^2}{32\pi^2}\,c_{qq} 
    + \sum_\ell\spac\frac{y_\ell^2}{32\pi^2}\,c_{\ell\ell}
    + \frac{27\alpha_s^2}{8\pi^2}\,C_F^{(3)}\spac\tilde c_{GG} 
    + \frac{9\alpha_2^2}{2\pi^2}\,C_F^{(2)}\spac\tilde c_{WW}
    + \frac{9\alpha_1^2}{8\pi^2} \left( {\cal Y}_L^2 + 3{\cal Y}_Q^2 \right)\!\spac \tilde c_{BB} 
    \notag\\
   &\approx \frac{3y_t^2}{32\pi^2}\,c_{tt} 
    + \frac{9\alpha_s^2}{2\pi^2}\,\tilde c_{GG} 
    + \frac{27\alpha_2^2}{8\pi^2}\,\tilde c_{WW}
    + \frac{3\alpha_1^2}{8\pi^2}\,\tilde c_{BB} \,, \notag\\
   \frac{d\spac\tilde c_{BB}}{d\ln\mu}
   &= \sum_i \left[ \frac{17 y_{u_i}^2}{96\pi^2}\,c_{u_i u_i}
    + \frac{5 y_{d_i}^2}{96\pi^2}\,c_{d_i d_i} + \frac{5 y_{e_i}^2}{32\pi^2}\,c_{e_i e_i} \right] \notag\\
   &\quad\mbox{}+ \frac{33\alpha_s^2}{8\pi^2}\,C_F^{(3)}\spac\tilde c_{GG}
    + \frac{3\alpha_2^2}{2\pi^2}\,C_F^{(2)}\spac\tilde c_{WW}
    + \frac{3\alpha_1^2}{8\pi^2} \left( 8{\cal Y}_u^2 + 2{\cal Y}_d^2 + {\cal Y}_Q^2 
    + 6{\cal Y}_e^2 + 3{\cal Y}_L^2 \right) \tilde c_{BB} \notag\\
   &\approx \frac{17y_t^2}{96\pi^2}\,c_{tt} 
    + \frac{11\alpha_s^2}{2\pi^2}\,\tilde c_{GG}
    + \frac{9\alpha_2^2}{8\pi^2}\,\tilde c_{WW}
    + \frac{95\alpha_1^2}{24\pi^2}\,\tilde c_{BB} \,.
\end{align}
Here $q=u,d,s,c,b,t$ and $\ell=e,\mu,\tau$ run over the various fermion flavors of the SM. We have derived these equations using the evolution equations (\ref{cVVRGE}) and (\ref{RGEs}). 

It is a very good approximation to drop all Yukawa couplings other than $y_t$, in which case only the ALP--fermion coupling $c_{tt}$ enters on the right-hand side of the equations. The evolution equation for this quantity can be derived from (\ref{RGEs}). In the approximation where only the top Yukawa is kept, one finds
\begin{equation}\label{cttRGE}
   \frac{d\spac c_{tt}}{d\ln\mu}
   \approx \frac{9 y_t^2}{16\pi^2}\,c_{tt}
    + \frac{2\alpha_s^2}{\pi^2}\,\tilde c_{GG} 
    + \frac{9\alpha_2^2}{16\pi^2}\,\tilde c_{WW}
    + \frac{17\alpha_1^2}{48\pi^2}\,\tilde c_{BB} \,.
\end{equation}
Relations (\ref{RGEs2l}) and (\ref{cttRGE}) form a coupled system of equations, which can be solved to obtain the scale-dependent coefficients $c_{tt}(\mu)$ and $\tilde c_{VV}(\mu)$. The solutions, in the approximation needed in this work, have been given in (\ref{cVVevol}) and (\ref{cQcu2}).

\section{Evolution equations for the effective Lagrangian (\ref{Leffalt})}
\label{app:B}
\renewcommand{\theequation}{B.\arabic{equation}}
\setcounter{equation}{0}

The effective Lagrangian (\ref{Leffalt}) provides an alternative description of the ALP interactions with SM fields. Here we present the RG evolution equations for the coupling parameters $\tilde c_{VV}$ and $\tilde{\bm{Y}}_f$ in this Lagrangian. The evolution equations for the quantities $\tilde c_{VV}$ have already been given in (\ref{RGEs2l}). These parameters are not scale independent, in contrast to the couplings $c_{VV}$ appearing in the original Lagrangian (\ref{Leff}). This fact may seem puzzling at first sight, because the operators describing the ALP--boson interactions are the same in the two forms of the effective Lagrangian. However, under renormalization the ALP--boson operators mix with the derivative couplings of the ALP to fermions in (\ref{Leff}). In the Lagrangian (\ref{Leffalt}) these derivative couplings must be decomposed into linear combinations of the non-derivative ALP--fermion interactions and the ALP--boson interactions. This decomposition introduces a non-trivial scale evolution of the parameters $\tilde c_{VV}$.

We have derived these equations for the coupling matrices $\tilde{\bm{Y}}_f$ starting from (\ref{rela1}) and the evolution equations for the parameters $c_{VV}$ and $\bm{c}_F$ given in (\ref{cVVRGE}) and (\ref{RGEs}), as well as the well-known RG equations of the SM Yukawa matrices \cite{Cheng:1973nv,Machacek:1981ic}. We obtain 
\begin{align}
   \frac{d}{d\ln\mu}\,\bm{\tilde{Y}}_u
   &= \frac{1}{16\pi^2} \left( 2\spac\tilde{\bm{Y}}_u \bm{Y}_u^\dagger \bm{Y}_u
    + \frac52\spac\bm{Y}_u \bm{Y}_u^\dagger \tilde{\bm{Y}}_u
    - \frac32\spac\bm{Y}_d \bm{Y}_d^\dagger \tilde{\bm{Y}}_u
    - 2\spac\bm{Y}_d \tilde{\bm{Y}}_d^\dagger \bm{Y}_u
    - \tilde{\bm{Y}}_d \bm{Y}_d^\dagger \bm{Y}_u \right) \notag\\
   &\mbox{}- \tilde{\bm{Y}}_u \left( \frac{2\alpha_s}{\pi} + \frac{9\alpha_2}{16\pi}
    + \frac{17\alpha_1}{48\pi} - \frac{T}{16\pi^2} \right) \notag\\
   &\mbox{}+ i\spac\bm{Y}_u \left[ - \frac{X}{8\pi^2} 
    + \frac{3\alpha_s^2}{2\pi^2}\,C_F^{(3)}\spac\tilde c_{GG}
    + \frac{3\alpha_2^2}{4\pi^2}\,C_F^{(2)}\spac\tilde c_{WW}
    + \frac{3\alpha_1^2}{4\pi^2} \left( \mathcal{Y}_u^2 + \mathcal{Y}_Q^2 \right) \tilde c_{BB}
    \right] , \notag\\
   \frac{d}{d\ln\mu}\,\bm{\tilde{Y}}_d 
   &= \frac{1}{16\pi^2} \left( 2\spac\tilde{\bm{Y}}_d \bm{Y}_d^\dagger \bm{Y}_d
    + \frac52\spac\bm{Y}_d \bm{Y}_d^\dagger \tilde{\bm{Y}}_d
    - \frac32\spac\bm{Y}_u \bm{Y}_u^\dagger \tilde{\bm{Y}}_d
    - 2\spac\bm{Y}_u \tilde{\bm{Y}}_u^\dagger \bm{Y}_d
    - \tilde{\bm{Y}}_u \bm{Y}_u^\dagger \bm{Y}_d \right) \notag\\
   &\quad\mbox{}- \tilde{\bm{Y}}_d \left( \frac{2\alpha_s}{\pi} + \frac{9\alpha_2}{16\pi}
    + \frac{5\alpha_1}{48\pi} - \frac{T}{16\pi^2} \right) \notag\\
   &\quad\mbox{}+ i\spac\bm{Y}_d \left[ \frac{X}{8\pi^2} 
    + \frac{3\alpha_s^2}{2\pi^2}\,C_F^{(3)}\spac\tilde c_{GG}
    + \frac{3\alpha_2^2}{4\pi^2}\,C_F^{(2)}\spac\tilde c_{WW}
    + \frac{3\alpha_1^2}{4\pi^2} \left( \mathcal{Y}_d^2 + \mathcal{Y}_Q^2 \right) \tilde c_{BB}
    \right] , \notag\\
   \frac{d}{d\ln\mu}\,\tilde{\bm{Y}}_e
   &= \frac{1}{16\pi^2} \left( 2\spac\tilde{\bm{Y}}_e \bm{Y}_e^\dagger \bm{Y}_e
    + \frac52\spac\bm{Y}_e \bm{Y}_e^\dagger \tilde{\bm{Y}}_e \right)
    - \tilde{\bm{Y}}_e \left( \frac{9\alpha_2}{16\pi} + \frac{15\alpha_1}{16\pi}
    - \frac{T}{16\pi^2} \right) \notag\\
   &\quad\mbox{}+ i\spac\bm{Y}_e \left[ \frac{X}{8\pi^2} 
    + \frac{3\alpha_2^2}{4\pi^2}\,C_F^{(2)}\spac\tilde c_{WW}
    + \frac{3\alpha_1^2}{4\pi^2} \left( \mathcal{Y}_e^2 + \mathcal{Y}_L^2 \right) \tilde c_{BB}
    \right] , 
\end{align}
with
\begin{equation}
   T = \text{Tr}\big( 3\bm{Y}_u^\dagger \bm{Y}_u + 3\bm{Y}_d^\dagger \bm{Y}_d
    + \bm{Y}_e^\dagger \bm{Y}_e \big) \,,
\end{equation}
and $X$ as given in (\ref{Xdef}). 

\end{appendix}

\end{document}